\documentclass[12pt]{article}
\pdfoutput=1
\usepackage{latexsym}
\usepackage{epsfig,amssymb,euscript}
\usepackage{amsmath} 
\usepackage{braket}
\usepackage{mathtools}
\usepackage{bbm,slashed}
\usepackage{array,calc,epsfig}
\usepackage{bbold}
\usepackage{color}
\usepackage{ulem}
\usepackage{cite}
\usepackage{graphicx}
\usepackage{hyperref}

\catcode`@=12

\numberwithin{equation}{section}

\begin{document}

\begin{titlepage}

\bigskip

\begin{center}
{\Large 
{\bf
Vacuum structure of large $\boldsymbol N$ QCD$_3$ from holography

%
}}
\end{center}

\bigskip
\begin{center}
{\large
Riccardo Argurio$^{1}$, Adi Armoni$^{2}$, Matteo Bertolini$^{3,4}$,\\
Francesco Mignosa$^{3}$, Pierluigi Niro$^{1,5}$}
\end{center}

\renewcommand{\thefootnote}{\arabic{footnote}}

\begin{center}
\vspace{0.2cm}
$^1$ {Physique Th\'eorique et Math\'ematique and International Solvay Institutes, \\ Universit\'e Libre de Bruxelles; C.P. 231, 1050 Brussels, Belgium\\}
$^2$ {Department of Physics, College of Science, Swansea University, SA2 8PP, UK\\}
$^3$ {SISSA and INFN - Via Bonomea 265; I 34136 Trieste, Italy\\}
$^4$ {ICTP - Strada Costiera 11; I 34014 Trieste, Italy\\}
$^5$ {Theoretische Natuurkunde, Vrije Universiteit Brussel;\\ Pleinlaan 2, 1050 Brussels, Belgium\\}
\vskip 5pt
{\texttt{rargurio@ulb.ac.be, a.armoni@swansea.ac.uk, bertmat@sissa.it, \\ fmignosa@sissa.it, pierluigi.niro@ulb.ac.be}}

\end{center}

\bigskip

\vskip 5pt
\noindent
\begin{center} {\bf Abstract} \end{center}
\noindent
We study the vacuum structure of three-dimensional $SU(N)$ gauge theory coupled to a Chern-Simons term at level $k$ and to $F$ fundamental Dirac fermions. We use a large $N$ holographic description based on a D3/D7 system in type IIB string theory compactified on a supersymmetry breaking circle. The multiple vacua of the theory and the transitions between them are nicely captured by the dual holographic background. The resulting phase diagram, which we derive both at leading and first subleading orders in the $1/N$ expansion, shows a rich structure where topological field theories, non-linear sigma models and first-order phase transitions appear.  

\end{titlepage}

\newpage
\tableofcontents


\section{Introduction}
\label{intro}
Establishing the properties of the vacuum of a strongly coupled quantum field theory is a notoriously difficult problem. For some classes of theories, typically gauge theories, that can be engineered in string theory, this problem can be translated into determining a vacuum configuration involving branes in a specific background. More precisely, the correspondence between gauge theories and string theory set-ups is particularly useful in the limit in which the string theory vacuum can be determined at the (semi-)classical level. This limit, which can be called the decoupling or holographic limit, involves taking the large $N$ and strong coupling limit on the gauge theory side. Hence it is perfectly suitable to determine the vacua of strongly coupled gauge theories. 

In this note, we focus on three-dimensional $SU(N)$ gauge theories with fundamental flavors, in the large $N$ limit. Since fermions are present, a Chern-Simons term will inevitably also be there, and indeed it will play a central role in the physics of the vacua of the theory. Our aim is to provide a simple string theory understanding of the various vacua that one finds as parameters are varied, and of the phase transitions between them.

Three-dimensional $SU(N)$ gauge theories with fundamental flavors, that we will call QCD$_3$, have been studied recently in several limits. The structure of the vacua depends on the following parameters, both discrete and continuous: the rank $N$, the number of flavors $F$, the Chern-Simons (CS) level $k$, and flavor masses $m$. In principle, flavors can have different masses, but we will always take them to be equal, unless otherwise stated, and call $m$ this common mass. The CS level can be defined in two equivalent ways: we can define a bare CS level $k_b$ by integrating out all fermions after giving them a large positive mass. Alternatively, we can define $k=k_b-F/2$ which has the property of flipping its sign under a time reversal transformation. We will actually see a natural string theory interpretation of both. Finally, we allow ourselves the slight abuse of language of calling $N$ the `rank' of $SU(N)$.

QCD$_3$ has been studied from the purely quantum field theory (QFT) side using many limits and conjectured dualities. The recent regain of interest started with \cite{Aharony:2011jz,Giombi:2011kc,Aharony:2012nh,GurAri:2012is,Aharony:2012ns,Jain:2013py,Jain:2013gza} where the large $N$, large $k$ limit was studied in the presence of a small number of flavors, and a conformal field theory (CFT) was conjectured to arise at vanishing flavor masses. Evidence were also presented for a dual description of such fixed point in terms of a theory with bosonic matter. This led, eventually, to the conjecture \cite{Aharony:2015mjs} that such boson/fermion duality (a.k.a.~bosonization) also holds at finite $N$, for $k\geq F/2$. It is morally a generalization of level/rank duality to CS gauge theories with matter, see \cite{Hsin:2016blu}. In this regime there is still a single phase transition at a critical value of $m$, but there is no direct handle to determine its order (but for the two cases where $F$ or $k$ are large enough, where it is known that the transition is second order \cite{Appelquist:1988sr,Appelquist:1989tc}). In \cite{Komargodski:2017keh} also the case $k<F/2$ was contemplated, and the phase diagram was conjectured to consist of three phases as $m$ is varied, with a purely quantum phase at small $m$ where the flavor symmetry $U(F)$ is spontaneously broken and the low-energy physics is captured by the corresponding Grassmannian $\sigma$-model. Again, little can be said about the order of the two phase transitions. Finally, in \cite{Armoni:2019lgb} the large $N$ limit at finite $k$ and $F$ was studied. A somewhat surprising result was found: irrespective of whether $k<F/2$ or $k\geq F/2$, a total of $F+1$ different phases were found, with generically a coexistence of topological and $\sigma$-model sectors. Moreover, they are separated by phase transitions that can be determined to be first order, following reasonings similar to the ones of \cite{Coleman:1980mx,Ferretti:1992fga,Ferretti:1992fd}. Quite interestingly, one should then expect a multicritical point for a (large) value of the CS level $k$ in which the phase transitions merge into a single second-order phase transition \cite{ADK}.

Below, we will propose a string theory picture for QCD$_3$, or more precisely for a gauge theory that we believe reproduces the low-energy behavior of QCD$_3$.\footnote{A string theory realization of QCD$_3$ at finite $N$, giving rise to bosonization and symmetry breaking phases in terms of a magnetic Seiberg dual theory, was proposed in \cite{Armoni:2017jkl,Akhond:2019ued}.} It refines a proposal made in \cite{Rey:2008zz,Hong:2010sb}, where (probe) flavors were added on the non-supersymmetric holographic description of Yang-Mills in three dimensions (YM$_3$), see \cite{Witten:1998zw,Aharony:1999ti} (and \cite{Fujita:2009kw} for the addition of a CS term in the set-up). In the body of the paper we will describe the technical details of our proposal, and some results concerning the phase diagram. Here, we want to outline by simple pictorial arguments how string theory helps us to find the different vacua of QCD$_3$, giving also evidence for the phase transitions being first order in the limit we are considering, which is the large $N$ limit at finite $k$ and $F$, i.e.~the same regime of the QFT analysis of \cite{Armoni:2019lgb}.

\subsection{A sketch of the brane construction for the QCD$_3$ vacua}
Our strategy for building a string theory configuration reproducing the physics of QCD$_3$ in the large $N$ limit with a finite CS level and a finite number of flavors will be to start from the string theory realization of YM$_3$ and then add both a CS term and flavors in a probe approximation. That means that YM$_3$ will be described by a curved background generated by the backreaction of some branes, while the other ingredients, flavors and CS levels, will be provided by other, non-backreacted branes (this is also called quenched approximation in holography).

$SU(N)$ YM$_3$ can be realized holographically as follows (see \cite{Witten:1998zw,Aharony:1999ti}). We take $N$ D3-branes of type IIB string theory, along the directions $x^0 \cdots x^3$, and thus orthogonal to $x^4\cdots x^9$. We further take $x^3$ to be compact with period $2\pi/M_{KK}$, and impose anti-periodic boundary conditions for the fermions in order to lift their zero-modes. Since quantum corrections lift scalar masses, we are left with pure $SU(N)$ YM$_3$ at energy scales below $M_{KK}$. We expect the backreacted geometry to encode the large $N$ regime of this gauge theory. Such geometry is given by the Euclidean black hole in $AdS_5\times S^5$. We will write its metric later on, here it is sufficient to state the topology of the resulting spacetime. If we call $r$ the radius of the space spanned by $x^4\cdots x^9$, the resulting geometry is (a warped version of)
\begin{equation}
\mathbb{R}^{1,2}\times {\cal C}_{r,x^3}\times S^5\ ,
\end{equation}
where $\mathbb{R}^{1,2}$ is spanned by $x^0\cdots x^2$ and ${\cal C}_{r,x^3}$ is the `cigar' geometry, which looks like a cylinder for large $r$ (as in the non-backreacted geometry), but ends smoothly at a finite value of $r$ (the tip of the cigar) and hence has the topology of a disk. It is this smoothness that encodes geometrically the confining property of the vacuum of YM$_3$. 

A CS level can be engineered following \cite{Fujita:2009kw}. We wrap D7-branes on the $S^5$ at the tip of the cigar, in such a way that their remaining three spacetime dimensions are along $\mathbb{R}^{1,2}$. As we will show later, before backreaction it can be seen that these D7-branes act like interfaces on the world-volume of the D3-branes. Moreover, they produce a monodromy for the RR field $C_0$. Because of the coupling involving $C_0$ in the Wess-Zumino term of the action of the D3s, this produces precisely a level $k$ CS term on the interfaces for the $SU(N)$ gauge fields, if there are $k$ D7-branes. Note that in the backreacted geometry there is a $U(k)$ gauge theory on the D7-branes, and moreover the five-form RR flux on the $S^5$ gives a CS level $-N$ to this effective 3d theory. Hence this set-up of D3s and D7s yields a holographic realization of the level/rank duality $SU(N)_k \leftrightarrow U(k)_{-N}$.

Finally, we add flavors as suggested in \cite{Hong:2010sb}. This is the 3d analog of what one does in the 4d version of this story, i.e.~the Sakai-Sugimoto model \cite{Sakai:2004cn}. One introduces flavor D7-branes, which in the present set-up are extended along $\mathbb{R}^{1,2}$ and the five space coordinates $x^4\cdots x^8$, and provide the matter fermions along the intersection with the D3-branes. After backreaction of the D3s there are two important differences with respect to the Sakai-Sugimoto model. First, the D7-branes have an extra orthogonal direction in which to go (in other words, they wrap an $S^4\subset S^5$), in addition to the angular coordinate on the cigar, so that they do not need to go back as anti-D7 branes. This is related to the fact that there is no gauge anomaly in 3d, hence anti-D7 branes are not required in the first place. Second, again because the D7-branes have the possibility to move along $x^9$, they are allowed to feel the repulsive force from the D3-branes, which is due to the fact that they have six mutually orthogonal directions. After backreaction, this is translated into the flavor D7-branes being slightly repelled from the tip of the cigar. 

\begin{figure}[h]
\includegraphics[scale=0.45]{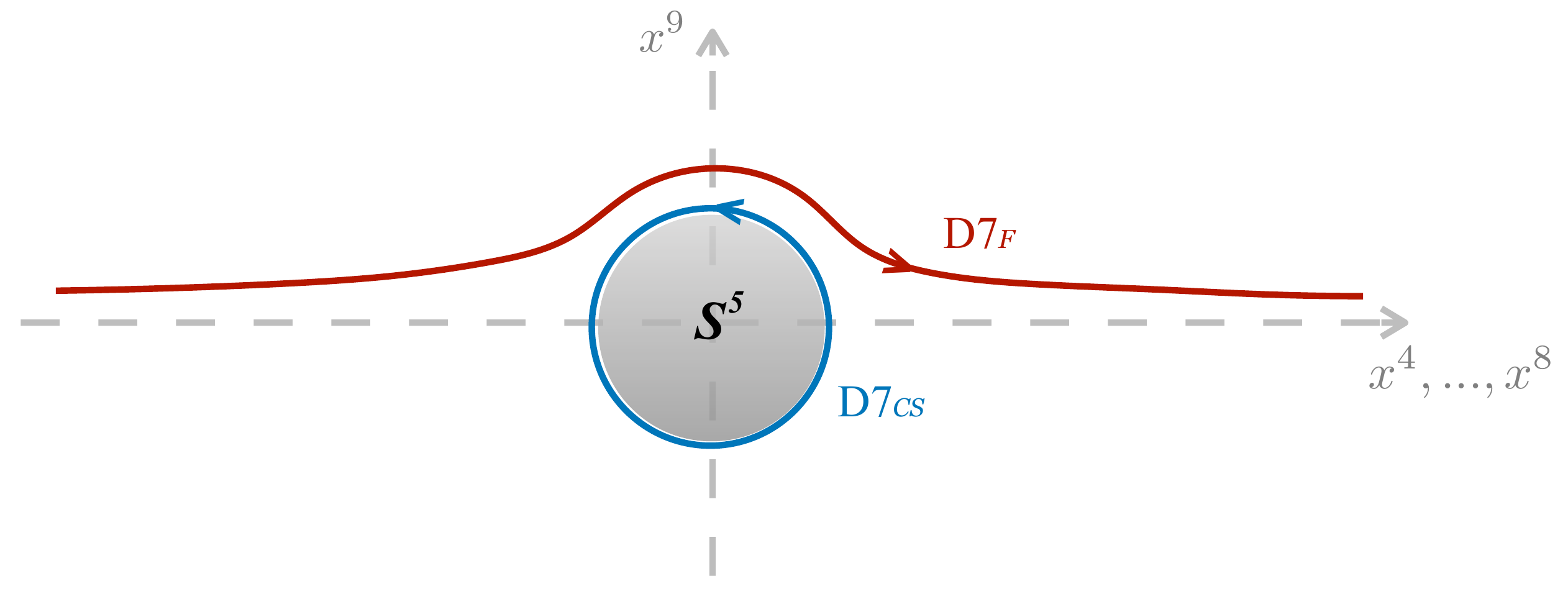}\centering
\centering
\caption{An artistic view of our brane set-up: the disk represents the $S^5$ at the tip of the cigar; the CS D7-branes wrap it, while the flavor D7-branes avoid it crossing the $x^9$ axis.}
\label{DKup.pdf}
\end{figure}

This is depicted in figure \ref{DKup.pdf}, which should be taken as an artistic rendering of the brane configuration, and similarly the following ones. Numerically exact graphs of the brane embeddings will be presented later. 

The embedding of the flavor D7-branes will depend on the boundary condition at infinity along the $x^4\cdots x^8$ directions. Since before backreaction and interactions are taken into account, the minimal distance between the D7-branes and D3-branes is given by the value of $x^9$ at which we place the D7s, we see that we will have massless fermions only if we set the D7s at $x^9=0$. Otherwise, the (bare) mass of the fermions will be proportional to the asymptotic value of $x^9$ of the D7-brane embedding.\footnote{This is another difference with respect to the Sakai-Sugimoto model, where it is notoriously subtle to introduce a mass for the fermions.} Note that this gives fermions a mass whose sign is flipped when the parity transformation is implemented by reversing the sign of $x^9$. In figure \ref{DKup.pdf} we have taken a positive value for $x^9$ at infinity, and we henceforth associate it to a positive fermion mass. 

\begin{figure}[h]
\includegraphics[scale=0.37]{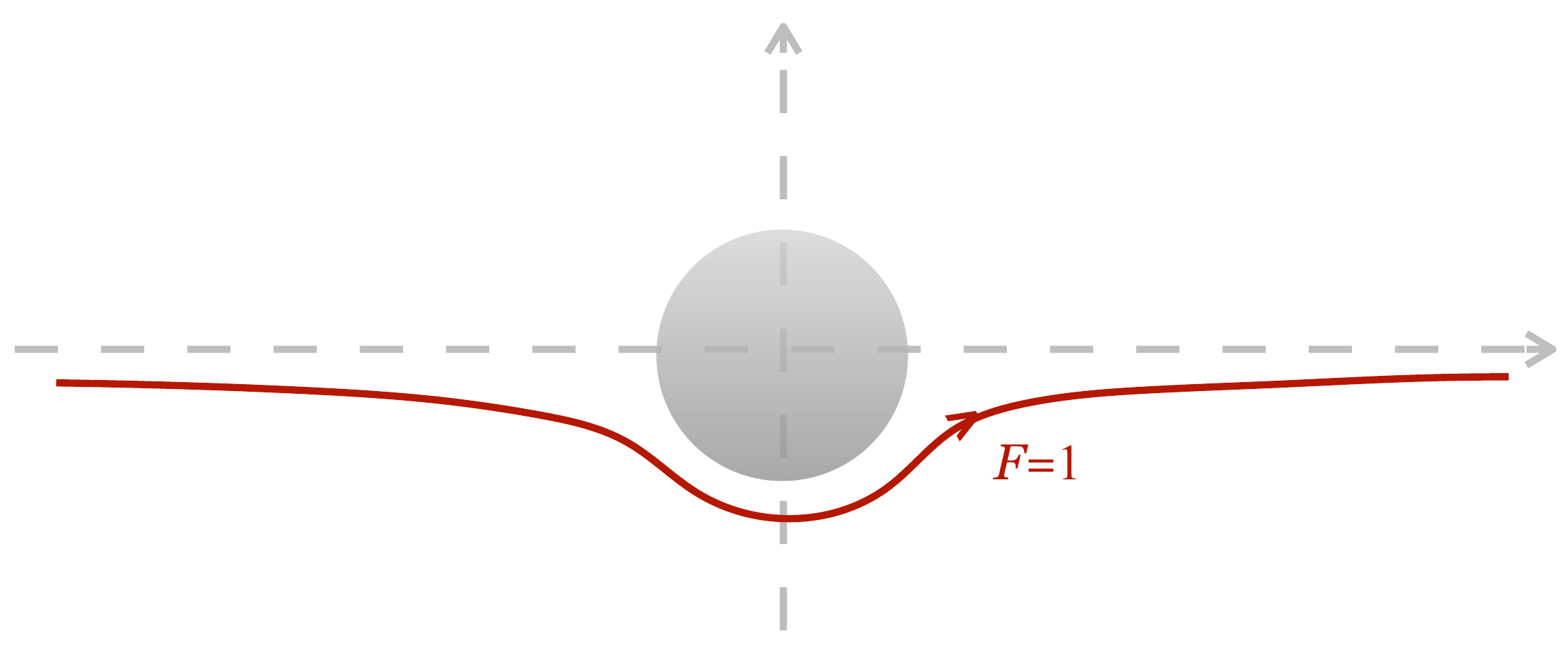}\centering
\centering
\caption{One flavor with negative mass and no CS branes.}
\label{DKdown.pdf}
\end{figure}

Let us now consider a configuration with one negative mass flavor and no CS branes, figure \ref{DKdown.pdf}. The embedding with minimal energy goes below the `disk.'

\begin{figure}[h]
\includegraphics[scale=0.33]{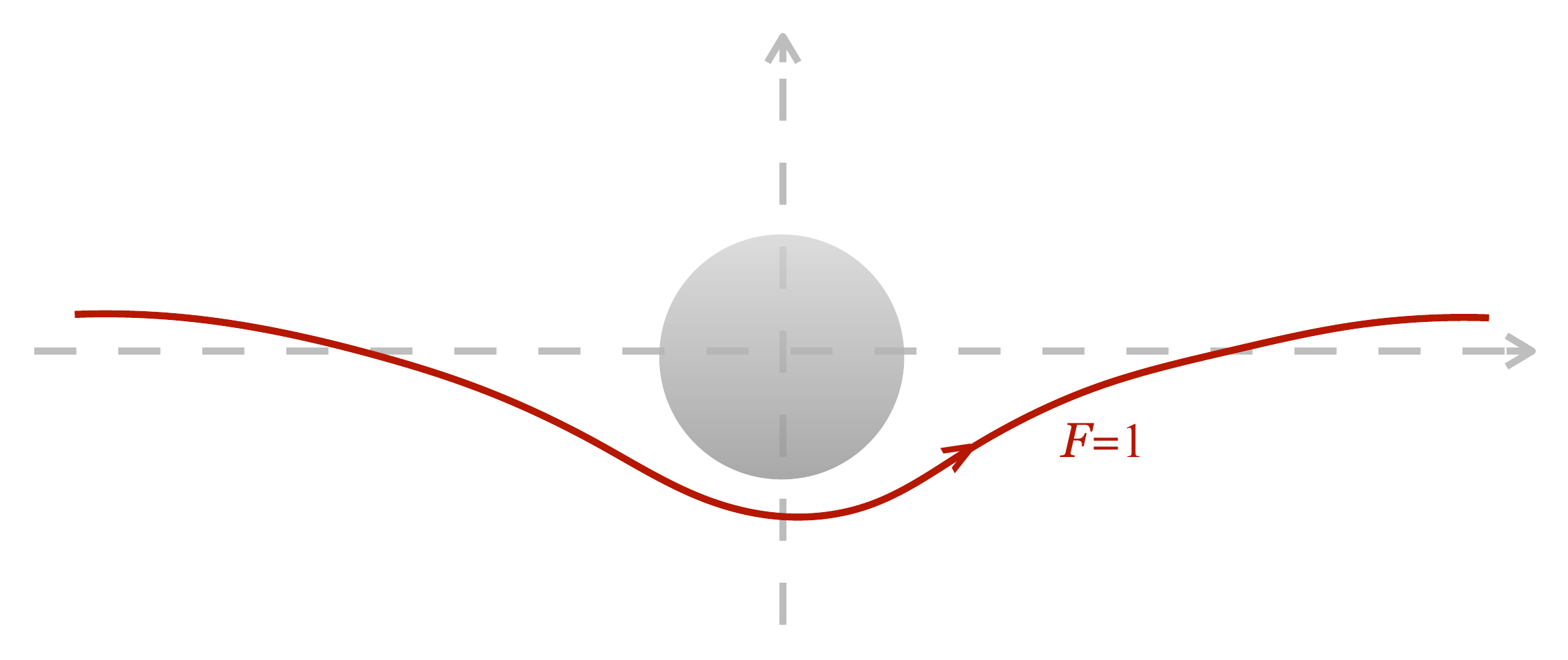}
\includegraphics[scale=0.33]{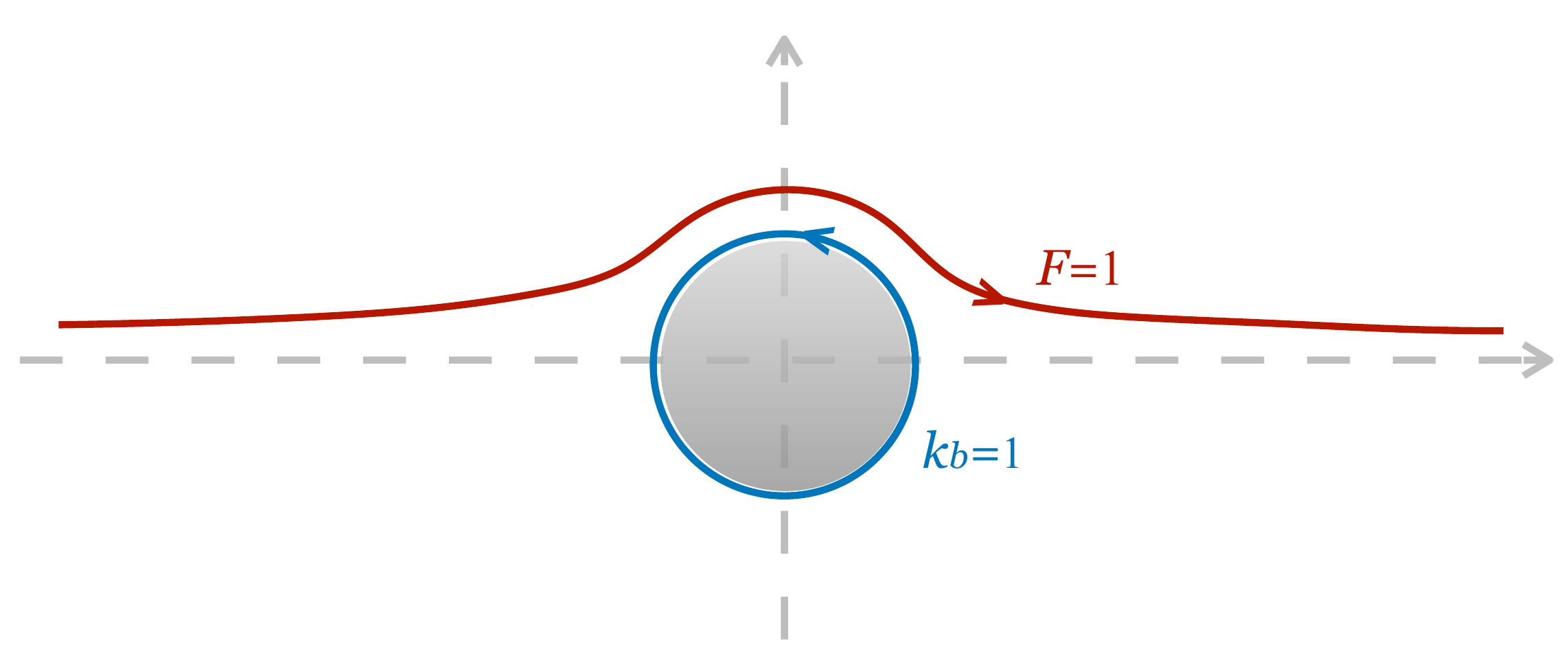}
\centering
\caption{The two embeddings for positive mass: the non-minimal without CS branes, and the minimal with one CS brane.}
\label{figcompet}
\end{figure}

If we now slowly bring the mass to positive values, i.e.~bring the asymptotic value of the embedding to positive values of $x^9$, the embedding will be deformed to a non-minimal one, still passing below the disk, as shown in figure \ref{figcompet} (left). However, there is another embedding with the same asymptotic value for $x^9$ and, importantly, the same D7 charge around the tip of the cigar: it is a minimal embedding going above the disk, accompanied by a CS brane wrapping the disk counterclockwise (same figure, right).

As we will discuss in detail, there is a critical bare mass above which the energetics will favor the figure on the right. Hence we are witnessing a transition between a theory whose vacuum is trivial, i.e.~$SU(N)_0$, for negative masses, to a theory with a topologically non-trivial vacuum, i.e.~$SU(N)_1$, for positive masses. This is nothing else but the two phases of the theory denoted as $SU(N)_{1/2}$ with one Dirac fermion in the fundamental.

\begin{figure}[h]
\includegraphics[scale=0.37]{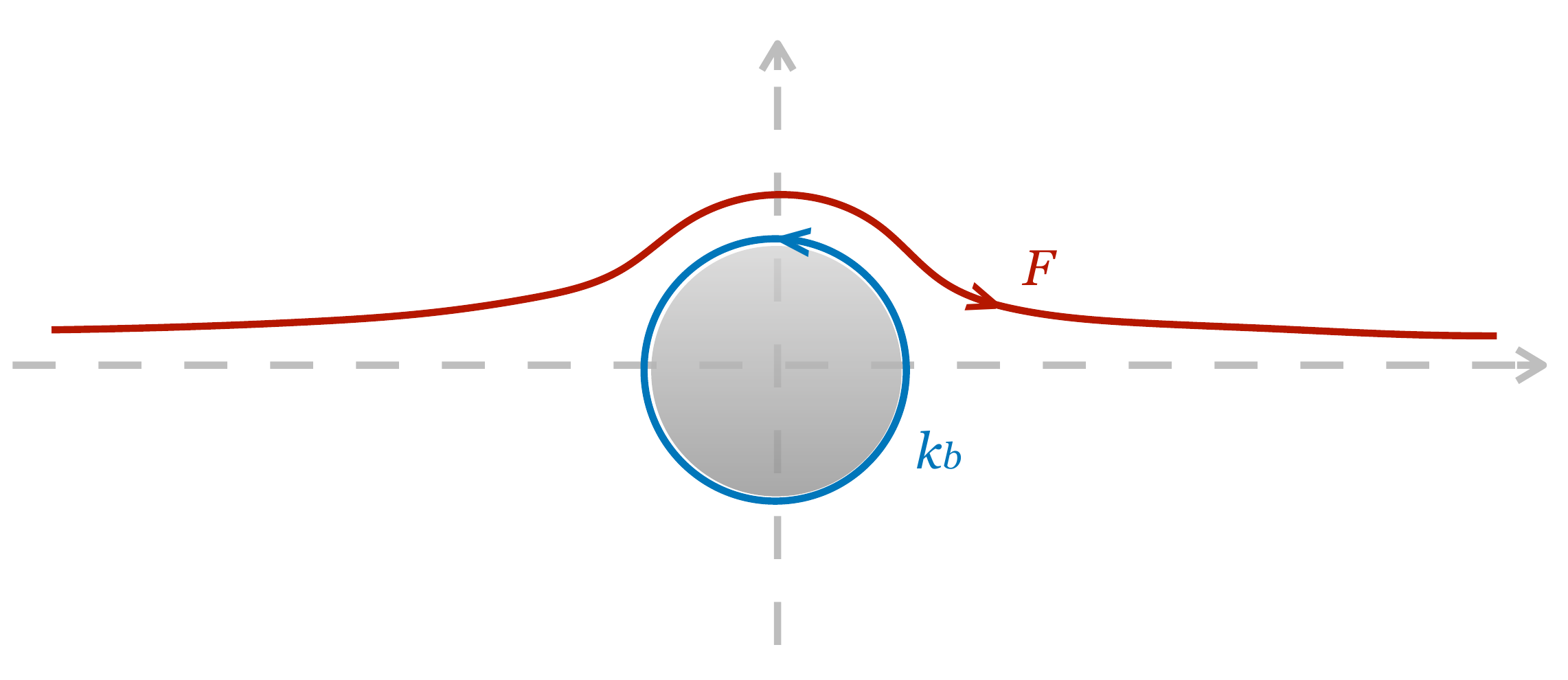}\centering
\centering
\caption{$F$ flavors with positive mass and a bare CS level given by $k_b$ (chosen to be positive in the figure).}
\label{DKupFkb}
\end{figure}

Let us now take a generic situation, with $F$ flavor D7-branes. In order to implement a well-defined CS level, we specify a bare CS level $k_b$, which is an integer, by wrapping $k_b$ D7-branes on the $S^5$ disk, counterclockwise if $k_b>0$ (and clockwise if $k_b<0$), when the flavors have a positive mass, i.e.~when flavor D7s are above the disk (figure \ref{DKupFkb}). 

The gauge theory on the CS D7-branes is $U(k_b)= U(k+F/2)$, where we used the relation between $k_b$ and $k$ in a theory with $F$ flavors, $k=k_b- F/2$. Its level is the same as in the YM$_3$ case, i.e.~$-N$. Hence for large positive mass, the vacuum is a topological theory, which by level/rank duality can be denoted by $SU(N)_{k+F/2}$, as in \cite{Aharony:2015mjs}. Note that the flavor D7-branes also carry a $U(F)$ gauge group, but since they extend infinitely in a direction orthogonal to $\mathbb{R}^{1,2}$, their coupling vanishes and it then corresponds to the global flavor symmetry of QCD$_3$.

Now bring all $F$ flavor branes below the disk, by tuning their mass to be large and negative. The minimal energy embedding will now pass below the disk, but in the rearranging process, or in other words, to conserve the D7 charge around the angular variable of the cigar, the number of CS D7-branes has to decrease by $F$ units. 

If $k_b\geq F$ (i.e.~$k\geq F/2$), we are left with $k_b-F$ CS branes, leading to a topological phase with $U(k_b-F)_{-N} = U(k-F/2)_{-N}$, level/rank dual to $SU(N)_{k-F/2}$, again as in \cite{Aharony:2015mjs}. 

\begin{figure}[h]
\includegraphics[scale=0.37]{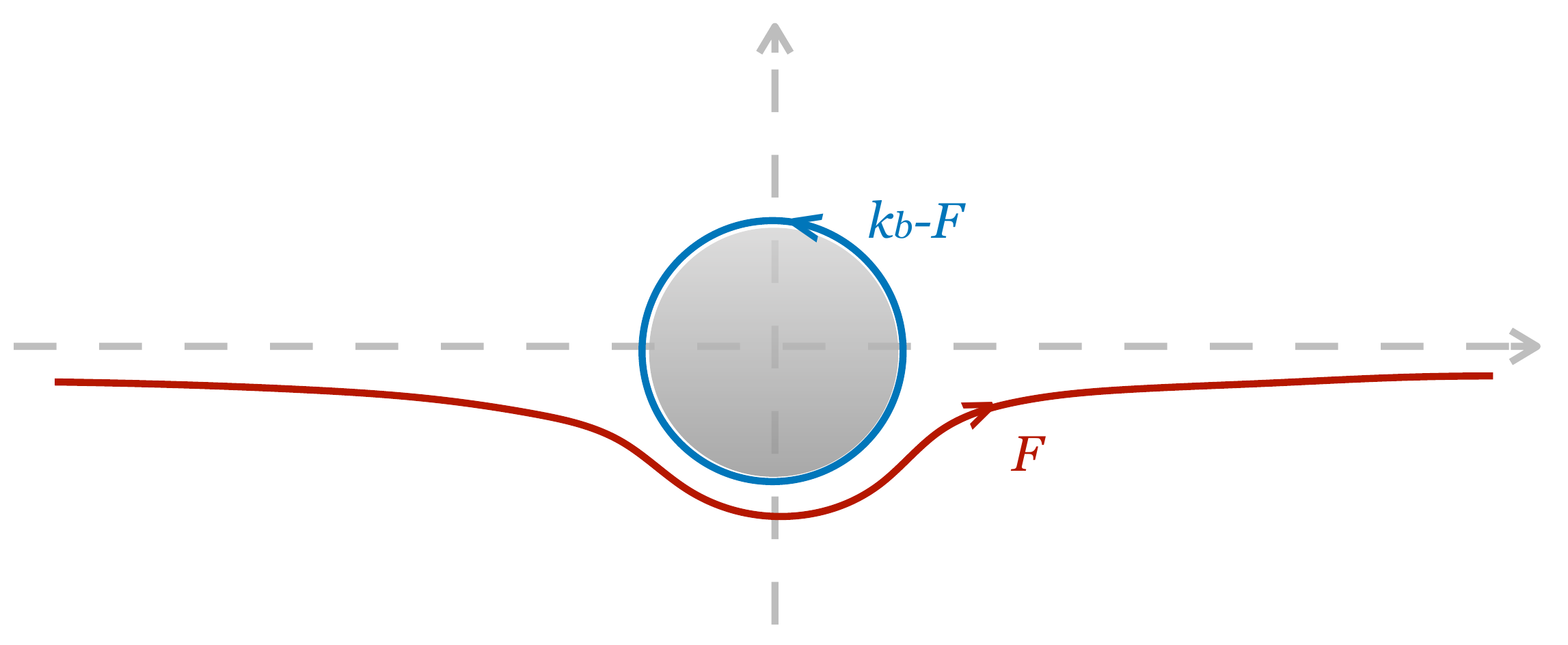}\centering
\centering
\caption{$F$ flavors with negative mass, and $k\geq F/2$.}
\label{DKduFkb}
\end{figure}

\begin{figure}[h]
\includegraphics[scale=0.37]{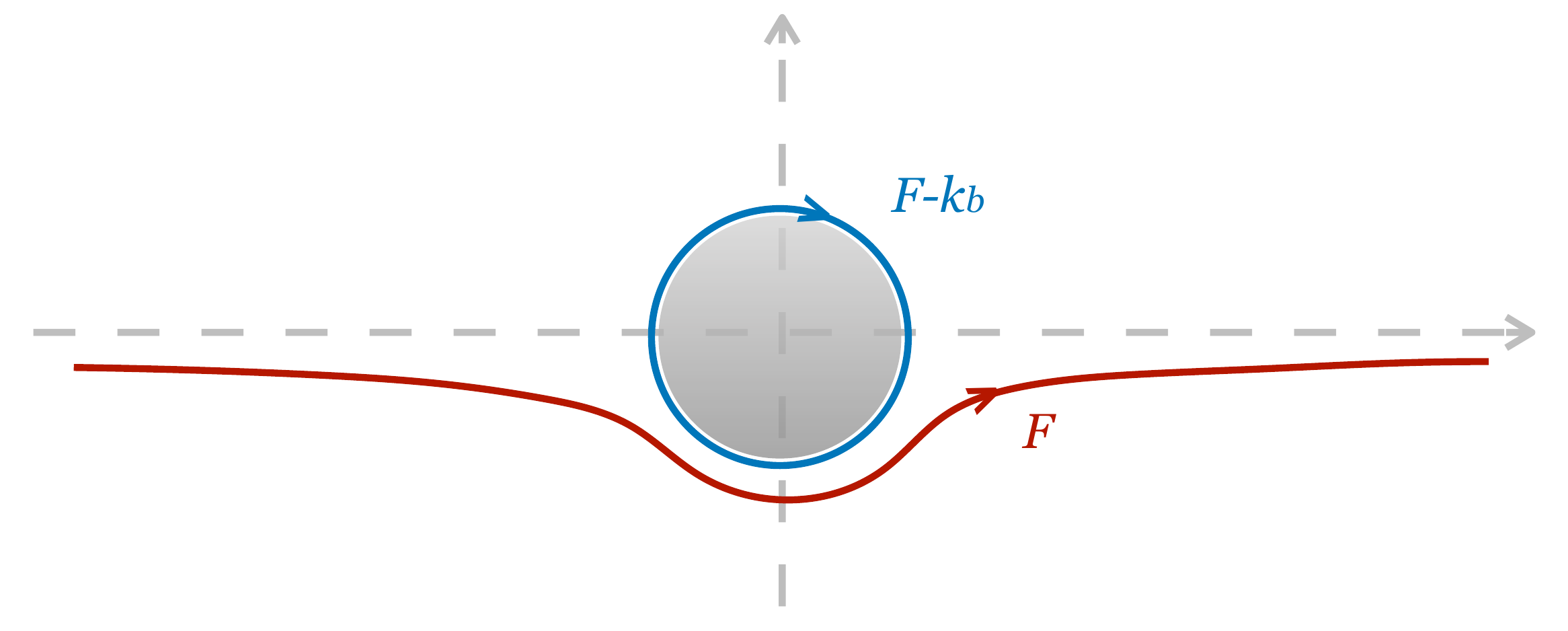}\centering
\centering
\caption{$F$ flavors with negative mass, and $k< F/2$.}
\label{DKdukbF}
\end{figure}

If $k_b< F$ (i.e.~$k< F/2$), we are left with $F-k_b$ clockwise CS branes, leading to a topological phase with $U(F-k_b)_{N} = U(F/2-k)_{N}$, in agreement with \cite{Komargodski:2017keh}. The level of the gauge theory flips sign since the D7-branes wrap the $S^5$ in the opposite way. The level/rank dual is $SU(N)_{k-F/2}$.
The two situations are depicted in figures \ref{DKduFkb} and \ref{DKdukbF}, respectively.

\begin{figure}[h]
\includegraphics[scale=0.37]{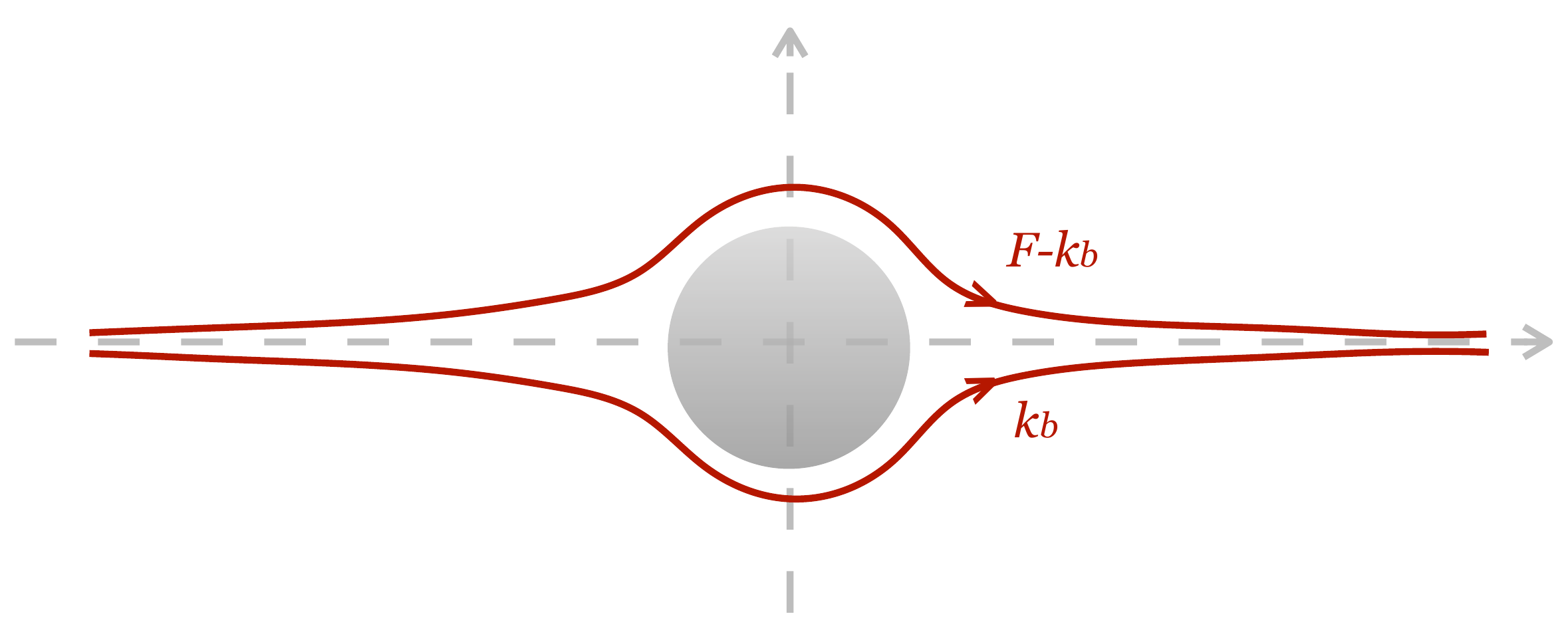}\centering
\centering
\caption{$F$ flavors with zero mass, splitting in such a way that no CS branes are left.}
\label{DKupdukb}
\end{figure}

We can now ask what happens for small masses, i.e.~for embeddings that asymptote to a small value of $x^9$. Take for instance the embedding that asymptotes to $x^9=0$, equivalent to a vanishing bare mass for the fermions. Obviously, the embeddings going above or below the disk have the same energy, since they are actually related by flipping the sign of $x^9$, i.e.~by a parity transformation. Since wrapping CS D7-branes on the $S^5$ costs some energy, we are then to conclude that the most favorable embedding is the one with the fewest CS branes. When $k_b<F$, this means that the true vacuum should not contain any CS brane at all, i.e.~there should not be any topological theory in the IR, as depicted in figure \ref{DKupdukb}. 

However, the IR theory is not empty, since the flavors have to split into $F-k_b$ above and $k_b$ below. Hence the flavor symmetry is spontaneously broken as
\begin{equation}
U(F)\to U(F-k_b)\times U(k_b) = U(F/2-k)\times U(F/2+k)\ ,
\end{equation}
and a $\sigma$-model parametrizing the Grassmannian $\mathrm{Gr}(F/2-k,F)$ arises from the corresponding Goldstone bosons. 

It is now manifest that there has to be a critical asymptotic value of $x^9$ such that the two configurations in figure \ref{figcompet} are isoenergetic. As we will see, this will happen for some positive value of the mass, $m^*$. 

\begin{figure}[h]
\includegraphics[scale=0.37]{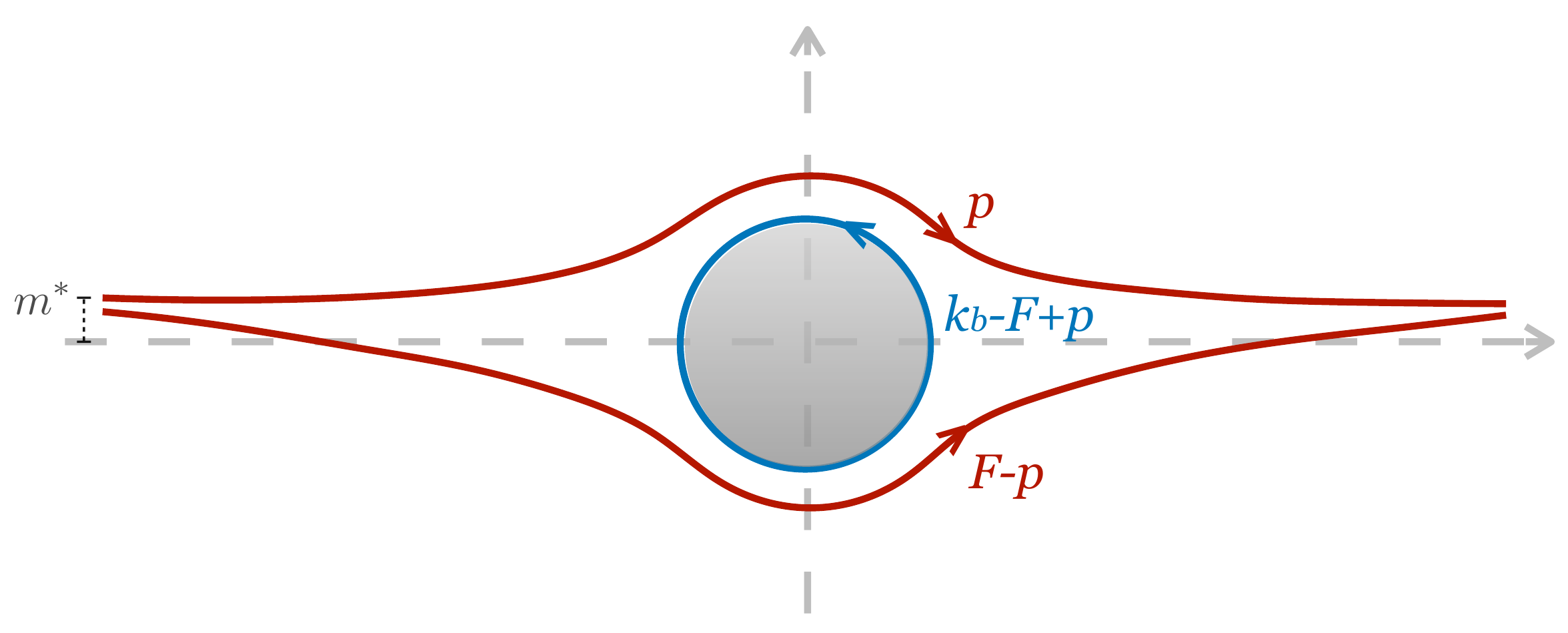}\centering
\centering
\caption{$F$ flavors at critical mass $m^*$, splitting in $F+1$ different ways, depending on the value of $p$.}
\label{DKupFp}
\end{figure}

Importantly, note that at such a critical mass there will be, generically, a degeneracy of more than two phases. Indeed, taking e.g.~$k_b\geq F$, all configurations represented in figure \ref{DKupFp} as $p$ is varied from 0 to $F$, have the same energy.

For every value of $p$ such that $0\leq p \leq F$, we have at low energies a $\sigma$-model on $\mathrm{Gr}(p,F)$ together with a topological CS theory $U(k_b-F+p)_{-N}= U(k-F/2+p)_{-N}$, for $k_b-F+p>0$. If $k_b-F+p<0$, the topological theory is $U(F-p-k_b)_{N}= U(F/2-p-k)_{N}$. In both cases the level/rank dual theory is $SU(N)_{k-F/2+p}$.
The Grassmannians accompanied by the topological theories are exactly the degenerate phases discussed for large $N$ QCD$_3$ in \cite{Armoni:2019lgb}, and we see them arising in this string theory construction in a very straightforward way.

Last but not least, we see pictorially that going from one phase to another requires some flavor branes to snap from above to below the disk. This clearly implies that the degenerate vacua are separated in field space, and therefore that the transitions are all first order.

The rest of the paper is organized as follows. In section 2 we provide details about the holographic model and we discuss the geometric properties of Chern-Simons and flavor probe branes. We use the holographic dictionary to extract information about the free energy and the fermion condensate on the field theory side. In section 3 we discuss the structure of brane configurations describing the different vacua of our holographic model and derive its phase diagram at leading order in the large $N$ expansion. We explicitly prove that there are multiple vacua and that the phase transitions in our model are first-order.  In section 4 we compute the $1/N$ corrections to the leading order phase diagram, and show how these modify its structure. Finally, in section 5, we focus on boson/fermion dualities and we show that our geometric set-up gives a simple understanding on how a dual bosonic description of QCD$_3$ arises at low energies. We conclude in section 6 with few more comments regarding the validity of the large $N$ expansion, brane backreaction and the existence of an IR fixed point.


\section{Holographic description of QCD$_3$}
\label{holo}

In this section we present the holographic set-up that we use to describe the large $N$ physics of QCD$_3$. We first address the case of pure Yang-Mills theory in presence of a Chern-Simons term, and then add flavor degrees of freedom.


\subsection{Yang-Mills theory with a Chern-Simons term}
\label{ymcs}

Three-dimensional $SU(N)$ Yang-Mills can be engineered by $N$ D3-branes wrapping a circle in one compactified direction, $x^3\simeq x^3+2\pi M_{KK}^{-1}$, with supersymmetry-breaking anti-periodic boundary conditions for the fermions \cite{Witten:1998zw}. In this way, the familiar four-dimensional $\mathcal{N}=4$ SYM theory reduces at low energies (below $M_{KK}$) to pure non-supersymmetric Yang-Mills theory in three dimensions. The dual type IIB supergravity background contains the metric $g$, the dilaton $\phi$ and the Ramond-Ramond five-form $F_5$
\begin{equation}
\begin{split}
&ds^2=\frac{r^2}{L^2}\left( \eta_{\mu\nu} dx^\mu dx^\nu + f(r) (dx^3)^2 \right) + \frac{L^2}{r^2 f(r)} dr^2 + L^2 d\Omega_5^2 \,, \\
&e^{\phi}=g_s \,, \qquad \frac{1}{(2\pi l_s)^4 }\int_{S^5} F_5 = N \,,
\end{split}
\label{background}
\end{equation}
where $\mu,\nu=0,1,2$ and
\begin{equation}
L^4=4\pi g_s N l_s^4 \,, \quad f(r)=1-\left(\frac{r_0}{r}\right)^4 \,, \quad r_0=\frac{M_{KK} L^2}{2} ~,
\end{equation}
with $l_s$ the string length and $g_s$ the string coupling. 
The geometry is given by a flat $\mathbb{R}^{1,2}$, a constant $S^5$ and a cigar-shaped $(r,x^3)$ subspace, where the holographic coordinate $r$ goes from $r_{UV}=\infty$ to the tip of the cigar $r=r_0$, where the geometry smoothly ends, thus giving rise to a mass gap and to confinement. In particular, two probe quarks at distance $d$ interact with a linear potential $V=\sigma d$, with \cite{Aharony:1999ti}
\begin{equation}
\sigma = \frac{\sqrt{g_sN}M^2_{KK}}{4\sqrt\pi} = \frac{\sqrt{\Lambda M^3_{KK}}}{4\sqrt{2\pi}} \,, 
\label{sigmadef}
\end{equation}
where $\Lambda$ is the scale of large $N$ QCD$_3$
\begin{equation}
\label{lambdaholo}
\Lambda \equiv g^2_{3d}N=\frac{g^2_{4d}}{2\pi M_{KK}^{-1}}N=\frac{4\pi g_s}{2\pi M_{KK}^{-1}}N=2 g_s N M_{KK} \,.
\end{equation}
The limit $L \gg l_s$, where the supergravity approximation is reliable, is equivalent to $g_s N \gg 1$, i.e.~$\Lambda \gg M_{KK}$, so there are spurious KK fields in this regime. In principle we should take the opposite limit $\Lambda \ll M_{KK}$, but this does not allow to use the supergravity approximation. We will comment on this opposite limit later. 

The D3 background reviewed above is dual to the strongly coupled regime of Yang-Mills theory. We now discuss the inclusion of a Chern-Simons term \cite{Fujita:2009kw}, ignoring flavor degrees of freedom for the time being. The D3-brane theory admits a coupling with the RR axion $C_0$
\begin{equation}
\label{C0coupl}
S_{C_0}=\frac{1}{4\pi} \int_{D3} C_0 \mbox{Tr} (F \wedge F) = - \frac{1}{4\pi} \int_{S^1} dC_0 \int_{\mathbb{R}^{1,2}} \omega_3(A) \,,
\end{equation}
where $\omega_3(A)$ is the Chern-Simons form in three dimensions and we have assumed that the gauge field on the D3-branes does not depend on $x^3$ and it does not have any components along $S^1$. If we choose (take $k_b$ non-negative)
\begin{equation}
dC_0(x^3) = - \frac{k_b}{2\pi M_{KK}^{-1} } dx^3 \,,
\label{rrpurecs}
\end{equation}
then we get the following term in the D3-brane action
\begin{equation}
\label{C0CS}
S_{C_0}= \frac{k_b}{4\pi} \int_{\mathbb{R}^{1,2}} \omega_3(A) \,,
\end{equation}
which is exactly a Chern-Simons term at level $k_b$. Clearly, we are neglecting the backreaction of the axion field on the D3 background, which is valid at leading order in $k_b/N$. Thus, the Minkowski part of the worldvolume of the D3-branes hosts an $SU(N)_{k_b}$ Yang-Mills theory at large $N$, with fixed $k_b$. At very low energies, this theory is believed to flow to a pure $SU(N)_{k_b}$ Chern-Simons theory.

At strong coupling, the stack of D3-branes is replaced by the cigar geometry, while the presence of a non-vanishing RR flux needs to be supported by a magnetic source for $C_0$. This is provided by $k_b$ probe D7-branes, which indeed couple magnetically to $C_0$. Being the number of branes an integer, this gives a holographic proof of the quantization of the Chern-Simons level. The CS branes wrap the $S^5$, share with the color D3-branes the three dimensions of Minkowski spacetime and are pointlike on the $(r,x^3)$ cigar. 

\begin{center}
\begin{tabular}{|c | c c c | c | c c|} 
\hline
 & 0 & 1 & 2 & 3 & r & $\Omega_5$ \\ [0.5ex] 
\hline
$N$ D3 & -- & -- & -- & -- & $\cdot$ & $\cdot$  \\
\hline
$k_b$ D7 & -- & -- & -- & $\cdot$ & $\cdot$ & -- \\
\hline
\end{tabular}
\end{center}

\begin{figure}[h]
\includegraphics[scale=0.35]{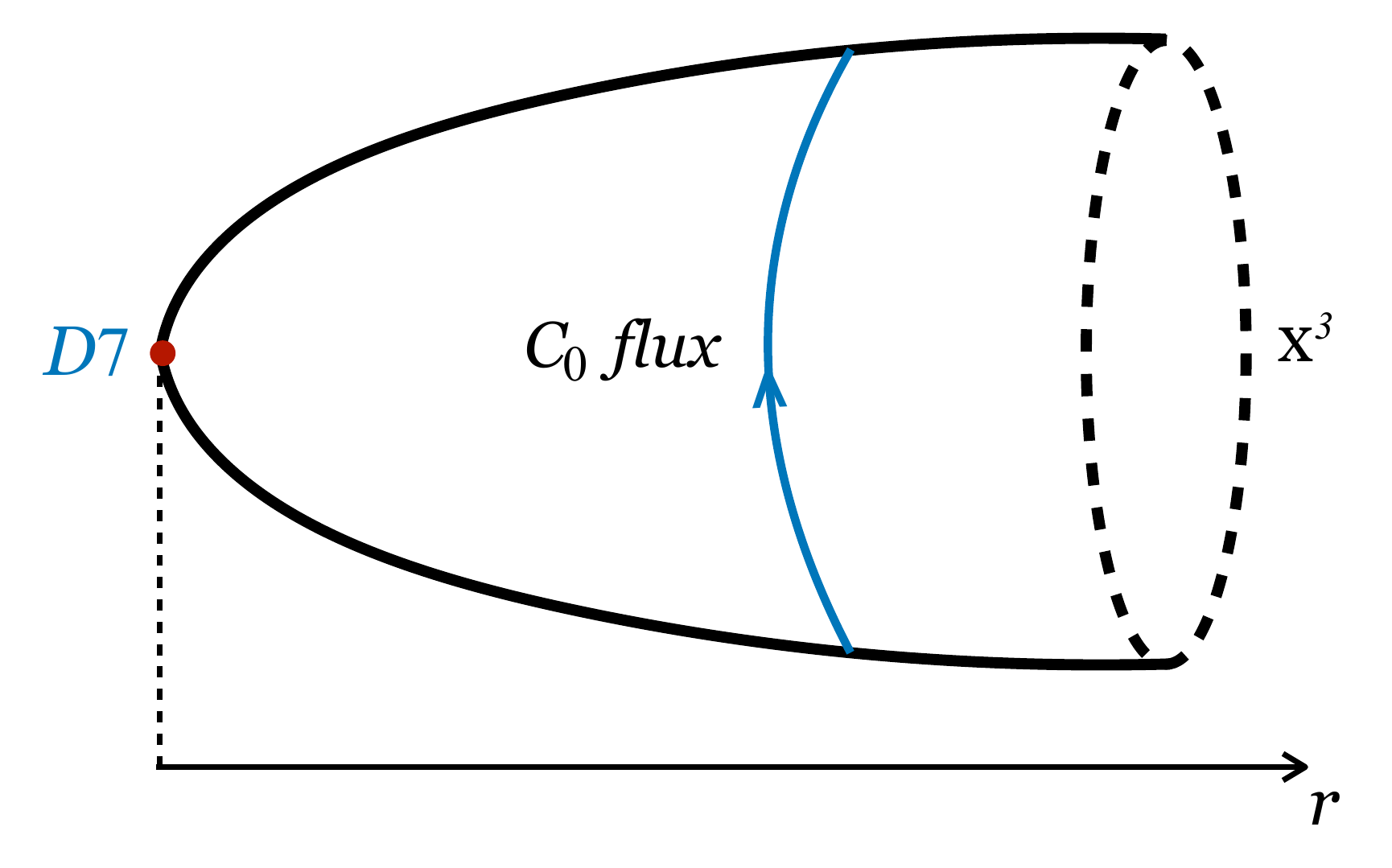}\centering
\centering
\caption{Chern-Simons D7-branes are located at the tip of the cigar and act as a source for the RR axion flux around $S^1$.}
\label{CSsig}
\end{figure}

Such D7-branes are located at its tip $r=r_0$, where the $x^3$ circle shrinks to a point, in order to minimize their energy density, see figure 
\ref{CSsig}. In this situation, the worldsheet of a string which is attached to a Wilson loop at the boundary can end on the D7-branes at the tip of the cigar.
This configuration is energetically favorite, as it has been explicitly computed in \cite{Fujita:2016gmu}, and signals a perimeter law for the Wilson loop. This shows, holographically, that in presence of a Chern-Simons term the theory does not confine. 

Following the standard holographic dictionary, the free energy density (in the three-dimensional sense) can be extracted from the on-shell value of the DBI action and reads, for a single CS brane
\begin{equation}
E_{CS} = -\frac{S_{D7}}{V_3} = T_{D7} V_5 L^2 r_0^3
\simeq N(g_s N)M^3_{KK}~,
\label{ECS}
\end{equation}
where $T_{D7}=(2\pi)^{-7}l^{-8}_sg_s^{-1}$ is the tension of the D7-brane, $V_3$ is the volume of three-dimensional Minkowski spacetime and $V_5$ is the volume of the unit five-sphere. In the strict large $N$ limit probe D-branes do not interact. Therefore, in this limit, the energy density of $k_b$ CS branes will be just $k_b E_{CS}$, which is linear in $N$. 

At energies below the $S^5$ inverse radius, the D7-brane theory reduces to a three-dimensional $U(k_b)$ gauge theory. Moreover, the presence of a background RR five-form flux induces a Chern-Simons term at level $-N$ from the corresponding Wess-Zumino term in the D7-brane action
\begin{equation}
S_{C_4}=\frac{1}{2 (2\pi)^5 l_s^4} \int_{D7} C_4 \wedge \mbox{Tr} (F \wedge F) = - \frac{1}{2 (2\pi)^5 l_s^4} \int_{S^5} F_5 \int_{\mathbb{R}^{1,2}} \omega_3 = - \frac{N}{4\pi}\int_{\mathbb{R}^{1,2}} \omega_3 \,.
\label{SC4}
\end{equation}
At very low energies, all excitations on the D7-branes decouple and we are left with a pure $U(k_b)_{-N}$ Chern-Simons theory. Thus, gauge/gravity duality in this set-up precisely reduces to the well-known level/rank duality $SU(N)_{k_b}\leftrightarrow U(k_b)_{-N}$ \cite{Fujita:2009kw}.

Note that if we take a negative $k_b$ the axion monodromy changes sign, meaning that we should put $|k_b|$ D7-branes with reversed orientation. This implies that there is a sign change in \eqref{SC4}, giving rise to the level/rank duality $SU(N)_{-|k_b|}\leftrightarrow U(|k_b|)_{N}$ at low energies, in agreement with QFT expectations. 


\subsection{Adding flavors in the holographic set-up}
\label{flavors}
As usual in AdS/CFT constructions, fundamental matter is added to the holographic set-up by introducing flavor branes in the background geometry \cite{Karch:2002sh}. In our case, we add $F$ copies of fundamental flavors by putting $F$ probe D7-branes, transverse to the compactified $x^3$ direction and spanning the Minkowski spacetime $\mathbb{R}^{1,2}$ and five of the six directions which are transverse to the D3-branes worldvolume. The leftover direction $x^9$ is transverse to both D3 and D7-branes. This configuration has 6 mixed Neumann-Dirichlet boundary conditions, thus breaking supersymmetry completely even in the case of a SUSY D3 background (i.e.~when $M_{KK}=0$). A bare mass for the flavors, which breaks parity in QCD$_3$, can be introduced by imposing a separation between color and flavor branes along $x^9$ at the UV boundary. 
 Indeed, the D7 worldvolume scalar corresponding to the $x^9$ direction couples to the fermionic mass operator. Consequently, the $x^9$ direction changes sign under the 3d parity transformation \cite{Rey:2008zz,Hong:2010sb}. According to the holographic dictionary, the profile of the flavor brane along $x^9$ is dual to the meson operator ${\bar\psi\psi}$ on the field theory side. 

\begin{center}
\begin{tabular}{|c | c c c | c | c c c c c c|} 
\hline
 & 0 & 1 & 2 & 3 & 4 & 5 & 6 & 7 & 8 & 9\\ [0.5ex] 
\hline
$N$ D3 & -- & -- & -- & -- & $\cdot$ & $\cdot$ &  $\cdot$ &  $\cdot$ &  $\cdot$ &  $\cdot$ \\
\hline
$F$ D7 & -- & -- & -- & $\cdot$ & -- & -- & -- & -- & -- & $\cdot$ \\
\hline
\end{tabular}
\end{center}
We will now analyze massless embeddings and then consider massive ones.


\subsubsection{Massless case}
\label{m=0}

It is convenient to describe the embedding in isotropic coordinates in the $x^4\cdots x^9$ directions, transverse to the D3-brane worldvolume. To achieve this, we first define a new radial coordinate $\rho$ such that
\begin{equation}
r(\rho)=\left(\rho^2+\frac{r^4_0}{4\rho^2}\right)^{1/2} \,.
\label{change}
\end{equation}
The ambiguity in inverting this relation is solved by choosing the branch $\rho^2 \geq r_0^2/2$, so that spacetime in the transverse directions does not extend towards the origin, but a five-sphere with radius $r_0/\sqrt{2}$ is excluded. In these coordinates the metric can be rewritten as (now $r=r(\rho)$)
\begin{equation}
ds^2=\frac{r^2}{L^2}\left( \eta_{\mu\nu} dx^\mu dx^\nu + f(r) (dx^3)^2 \right) + \frac{L^2}{\rho^2} \left( d\rho^2 + \rho^2 d\Omega_5^2 \right) \,.
\end{equation}
We now separate the six transverse coordinate in the $x^4\cdots x^8$ directions, which are part of the flavor branes worldvolume and for which we choose spherical coordinates $\lambda$ and $\Omega_4$ (with $\lambda\geq 0$), and the transverse 9 direction $u\in(-\infty,+\infty)$. The final form of the metric is
\begin{equation}
\label{bgD7f}
ds^2=\frac{r^2}{L^2}\left( \eta_{\mu\nu} dx^\mu dx^\nu + f(r) (dx^3)^2 \right) + \frac{L^2}{\rho^2} \left( d\lambda^2 + \lambda^2 d\Omega_4^2 + du^2 \right) \,,
\end{equation}
where $r=r(\rho)$ as in \eqref{change} and $\rho^2=\lambda^2+u^2 \geq r_0^2/2$. With this choice the D7-brane worldvolume is spanned by the eight coordinates $s=(x^\mu,\lambda,\Omega_4)$ and its embedding is described by $(x^3,u)=(x^3(s),u(s))$. We set $x^3$ to a constant, meaning that the D7-brane is localized on the circle, and by translational and rotational symmetry $u=u(\lambda)$. We have reduced the problem of finding the D7-brane embedding to the problem of finding the profile of a real function of a single real and positive variable. Recall that a parity transformation acts as $u(\lambda)\rightarrow -u(\lambda)$ and that, in the massless case, we have to impose the following boundary conditions ($\dot u \equiv du/d\lambda$ from now on)
\begin{equation}
\dot u (0) = 0 \,, \qquad u(\lambda_\infty)=0 \,,
\label{boundcond}
\end{equation}
where $\lambda_\infty$ is the location of the boundary, related to the UV cutoff on the field theory side. Now we are ready to compute the differential equation that $u(\lambda)$ should satisfy. First of all, the induced metric on the D7 takes the form
\begin{equation}
ds^2|_{D7}=\frac{r^2}{L^2}\eta_{\mu\nu} dx^\mu dx^\nu + \frac{L^2}{\rho^2} \left( (1+\dot u^2) d\lambda^2 + \lambda^2 d\Omega_4^2 \right) \,,
\end{equation}
so that the action for a single D7 is
\begin{equation}
S_{D7} = - \frac{1}{(2\pi)^7l_s^8} \int d^8s ~ e^{-\phi} \sqrt{-g|_{D7}} = -T_{D7} V_3 V_4 L^2 \int d\lambda \left(\rho^2+\frac{r^4_0}{4\rho^2}\right)^{3/2} \frac{\lambda^4}{\rho^5} \sqrt{1+\dot u^2} \,,
\label{action}
\end{equation}
where $T_{D7}=(2\pi)^{-7}l_s^{-8}g_s^{-1}$ is the D7-brane tension, $V_3$ is the volume of Minkowski spacetime and $V_4$ is the one of the unit four-sphere. The Euler-Lagrange equation of motion describing the D7-brane embedding is
\begin{equation}
\frac{d}{d\lambda} \left[ (r^4_0+4\rho^4)^{3/2} \frac{\lambda^4}{8\rho^8} \frac{\dot u}{\sqrt{1+\dot u^2}} \right] = - (r^4_0+\rho^4)(r^4_0+4\rho^4)^{1/2} \frac{\lambda^4 u}{\rho^{10}} \sqrt{1+\dot u^2} \,,
\label{fulleq}
\end{equation}
to be solved with boundary conditions \eqref{boundcond}. Few observations are in order.

\begin{itemize}

\item If $r_0=0$, then the equation of motion reads
\begin{equation}
\frac{d}{d\lambda} \left[\frac{\lambda^4 \dot u}{\rho^2 \sqrt{1+\dot u^2}} \right] = - \frac{2\lambda^4 u}{\rho^{4}} \sqrt{1+\dot u^2} \,.
\end{equation}
Being the right-hand side non-vanishing, it is easy to see that a constant profile for $u(\lambda)$ is a solution only if $u(\lambda)\equiv 0$.\footnote{As opposite to what happens for the D4/D6 system of \cite{Kruczenski:2003uq} and for the D3/D5 system of \cite{Jensen:2017xbs}, where any constant profile is a solution for $r_0=0$. This reflects the fact that even in the supersymmetric case $M_{KK}=0$ the D3/D7 system we consider is non-BPS.} The solution $u(\lambda)=0$ implies no symmetry breaking at all (being invariant under $u\rightarrow -u$), so we expect it to be unstable, as it was verified, for instance, in \cite{Kutasov:2011fr}.

\item An exact solution of \eqref{fulleq} is given by the profile which wraps half of the five-sphere and then sits at $u=0$ to $\lambda=\infty$, i.e.
\begin{equation}
u(\lambda)=
\begin{cases}
\pm \sqrt{r_0^2/2-\lambda^2}  ~ &\mbox{if} ~ 0\leq \lambda \leq r_0/\sqrt{2} \,, \\
0 ~ &\mbox{if} ~ \lambda \geq r_0/\sqrt{2} \,. \\ 
\end{cases}
\label{maxemb}
\end{equation}
There are two solutions, corresponding to the two signs in \eqref{maxemb}, one being a D7-brane wrapping the upper half-five-sphere and the other a D7-brane wrapping the lower one, as shown in figure \ref{maxembfig}. We will refer to these profiles as the maximal embeddings, since we will show that these solutions correspond to the ones having maximal energy, among all possible solutions to \eqref{fulleq}.

\begin{figure}[h]
\includegraphics[scale=0.28]{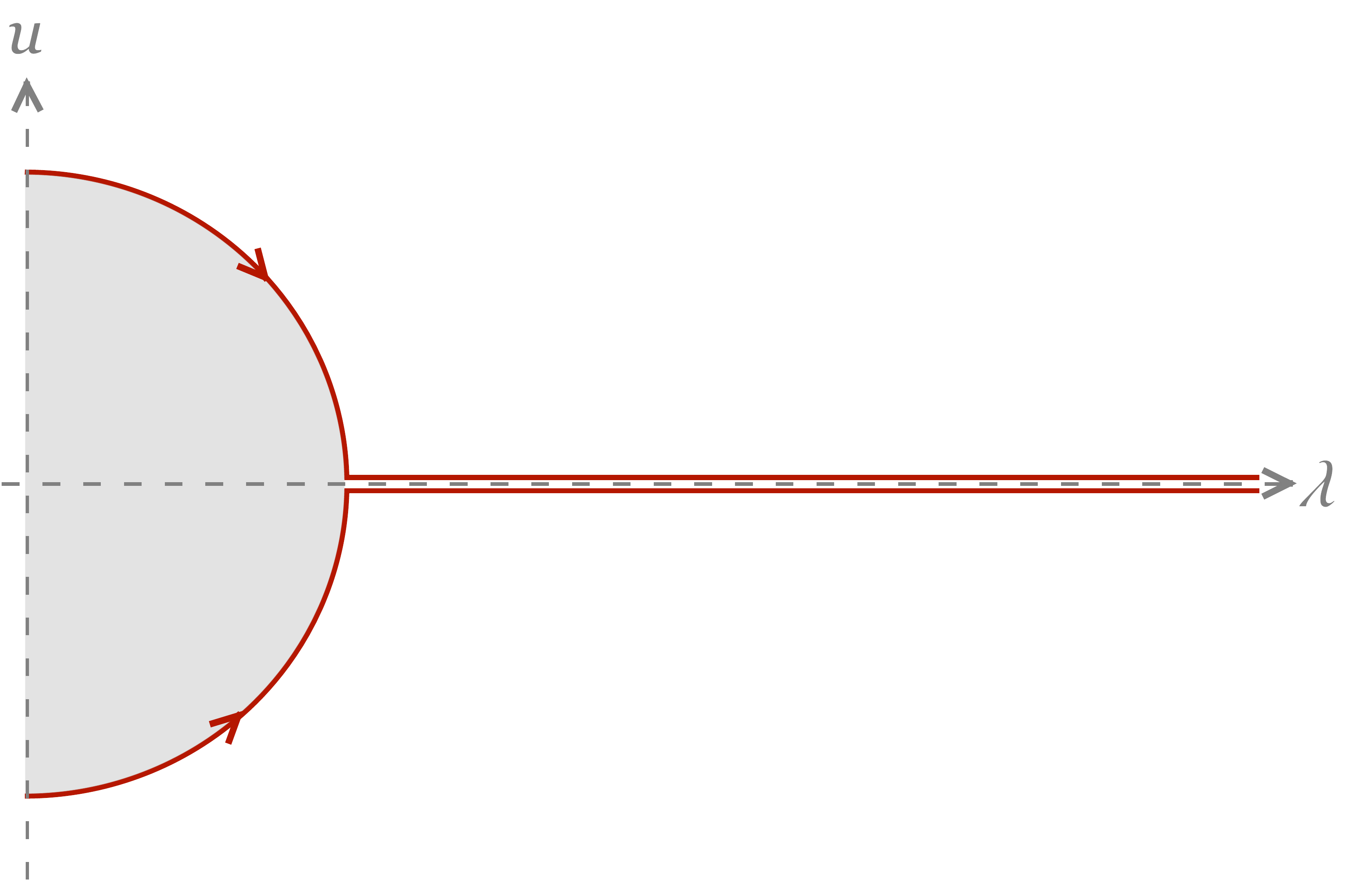}\centering
\centering
\caption{The two parity related maximal embeddings, corresponding to the highest energy solutions of the differential equation for the flavor brane profile.}
\label{maxembfig}
\end{figure}

\item In this choice of coordinates, the embedding $r=r_0$ representing the CS D7-branes discussed in section \ref{ymcs} takes the form
\begin{equation}
u^2(\lambda)=r_0^2/2-\lambda^2 \,,
\end{equation}
so that the full sphere is wrapped by the CS branes, which do not reach the UV boundary (and hence they do not introduce new degrees of freedom on the dual field theory side), see figure \ref{csbranefig}. We can choose an orientation for the CS branes, which are semicircles in the $(\lambda,u)$ plane. In the conventions where the flavor D7-branes are taken to be oriented from left to right, a positive (negative) $k_b$ is given by $|k_b|$ counterclockwise (clockwise) CS D7-branes. As we will explain in the next section, this is consistent with integrating out massive fermions in QCD$_3$.

\begin{figure}[h]
\includegraphics[scale=0.29]{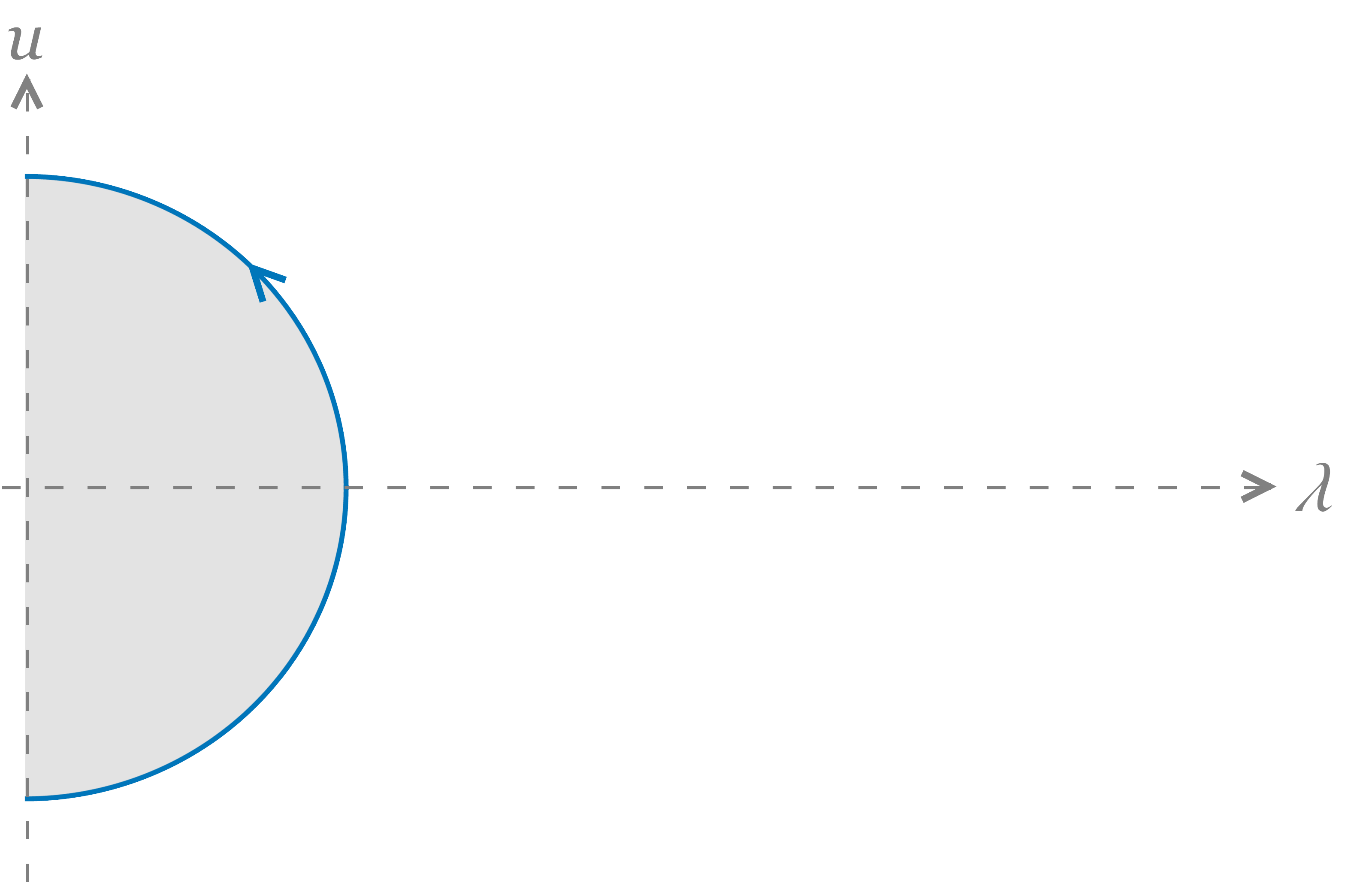}\centering
\centering
\caption{Chern-Simons branes are semicircles in the $(\lambda,u)$ plane. A positive level correponds to the counterclockwise orientation.}
\label{csbranefig}
\end{figure}

\item Most importantly, if $\tilde u(\lambda)$ is a solution, then also $- \tilde u(\lambda)$ is a solution (for this to hold it is crucial that the boundary condition $u(\lambda_\infty)=0$ is parity-invariant). These solutions are related by a 3d parity transformation and have the same energy, simply because \eqref{action} is parity-invariant. This is the first achievement of holography: the fact that the effective potential for the eigenvalues of $\braket{\bar\psi\psi}$ has two degenerate minima at opposite non-vanishing values is an assumption in \cite{Armoni:2019lgb}. Here instead, this is geometrically realized in a natural way.
\end{itemize}

Let us start discussing the solution in the asymptotic region where $\lambda$ is close to the cutoff.\footnote{An alternative regularization that does not require a cutoff consists in considering the full asymptotically flat D3-brane metric, and not only its near-horizon limit \cite{Kristjansen:2016rsc}. We prefer not to use this regularization since its holographic interpretation is less clear.} 
There $u(\lambda)$ is small and we will assume it to be slowly varying, so that the equation of motion becomes
\begin{equation}
\frac{d}{d\lambda}(\lambda^2 \dot u)=-2u \,.
\label{aseq}
\end{equation}
This equation is scale invariant, which reflects the absence of a scale in the theory in the far UV where $r_0$ is negligible (this QCD$_3$ UV-completion is $\mathcal{N}=4$ SYM with 3d defect fermions, as in \cite{Kutasov:2011fr}).  The characteristic polynomial has two complex conjugate roots $\alpha_\pm=(-1\pm i\sqrt{7})/2$. The appearance of complex roots is because this equation describes the propagation of a field whose mass is below the Breitenlohner-Freedman bound \cite{Breitenlohner:1982jf}, as already emphasized in \cite{Rey:2008zz,Kutasov:2011fr}.  The violation of the BF bound corresponds to the instability of the embedding defined in eq.~\eqref{maxemb}, which indeed does not represent the minimal energy configuration. This has been analyzed in detail in the case $r_0=0$ in \cite{Kutasov:2012uq}, where it was shown that only the minimal energy configurations are free of tachyon instabilities. As we will discuss later, the instability of the maximal embedding is related to the loss of conformality of the dual field theory.

The general form of the large $\lambda$ behavior of the solution is
\begin{equation}
u(\lambda) = \pm \sqrt{\frac{\mu^3}{\lambda}} \sin \left( \frac{\sqrt 7}{2} \log \frac{\lambda}{\mu} + \varphi \right) \,,
\label{generaluinf}
\end{equation}
where $\mu$ and $\varphi$ are the two integration constant, being $\mu$ a positive quantity with the dimension of a length. As we will see later, this scale is related to the scale of the fermion condensate. The two signs are related to the two possible parity-related choices.

Requiring that $u(\lambda_\infty)=0$, we can determine $\varphi$ and the asymptotic solution reads
\begin{equation}
u(\lambda) = \pm \sqrt{\frac{\mu^3}{\lambda}} \sin \left( \frac{\sqrt 7}{2} \log \frac{\lambda}{\lambda_\infty} \right) \,.
\label{uinf}
\end{equation}
We also get
\begin{equation}
\dot u(\lambda_\infty) = \pm \frac{\sqrt 7}{2} \left(\frac{\mu}{\lambda_\infty}\right)^{3/2} \,.
\label{asympexpansion}
\end{equation}
Near the cutoff the global signs of the $u$ and $\dot u$ are opposite: this means that the embedding with $\dot u(\lambda_\infty)>0$ ($\dot u(\lambda_\infty)<0$) approaches zero from negative (positive) values of $u$. The scale $\mu$ is fixed completely by the initial conditions at $\lambda=0$ and it can be thought to be of the order of $u(0)$. Also, as a consistency check, we found $\dot u$ to be always numerically small (although non-vanishing) close to $\lambda_\infty$, for any value of the cutoff.  

Let us now discuss the solution in the other asymptotic region, where $\lambda$ is small, $\dot u(\lambda) \sim 0$ and $\rho\sim u$. The behavior of $u(\lambda)$ for small values of $\lambda$ is given by
\begin{equation}
u(\lambda)=u_0-\frac{4(r^4_0+u_0^4)}{5u_0(r^4_0+4u_0^4)}\lambda^2 \,.
\label{u0}
\end{equation}
Note that the second derivative $\ddot u(0)$ has the opposite sign with respect to $u(0)=u_0$, meaning that the brane profile tends to bend towards the horizontal axis $u=0$. 

\begin{figure}[h]
\includegraphics[scale=0.28]{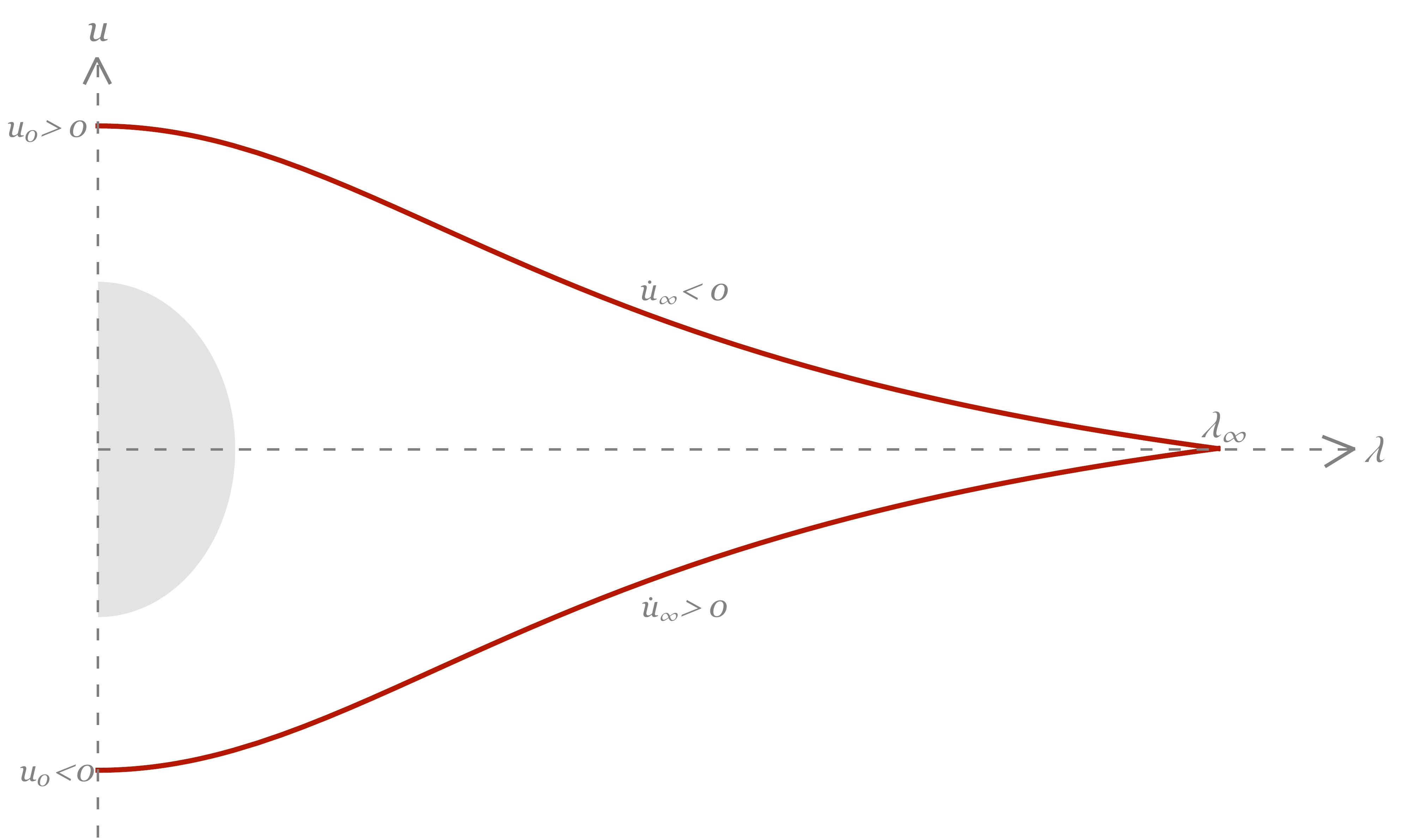}\centering
\centering
\caption{The two parity-related minimal embeddings in the massless case, numerical result with parameters $\mu=1.34 \, r_0$ and $\lambda_\infty=5.73 \,r_0$.}
\label{Masslessfig}
\end{figure}

The asymptotic expansions of $u(\lambda)$ for large and small $\lambda$ are given by \eqref{uinf} and \eqref{u0}, respectively. It is natural to make the following correspondence between the sign ambiguities of these formulas: if $u_0>0$ in \eqref{u0} then the minus sign should be chosen in \eqref{uinf}, and viceversa. This is confirmed by numerical analysis, as figure \ref{Masslessfig} shows. In this way, the D7-brane embedding interpolates from $u(0)=u_0$ to $u(\lambda_\infty)=0$ monotonically without crossing the $u=0$ axis. This zero-node embedding comes with two isoenergetic parity-related solutions, the `up' with $u_0>0, \dot u_\infty <0$ and the `down' with $u_0<0,\dot u_\infty>0$.

All other solutions of \eqref{fulleq} are given by multiple-node functions $u(\lambda)$ and it is easy to show numerically that the associated energy is an increasing monotonic function of the number of nodes. In particular, we can regard the maximal embeddings \eqref{maxemb} as the ones having the highest energy, since the constant behavior $u=0$ for large values of $\lambda$ can be seen as an  embedding with infinite number of nodes.

Let us now compute the energy density associated to the flavor branes. For the maximal embedding it can be computed analytically and it reads, up to terms suppressed by $(r_0/\lambda_\infty)^4$
\begin{equation}
E^{max}_{D7}=-\frac{1}{V_3}S^{max}_{D7} = T_{D7} V_4 L^2 \left(\frac{\lambda^3_\infty}{3}+b_{max}r_0^3 \right)\,,
\label{maxenergy}
\end{equation}
where
\begin{equation}
b_{max}= \frac{3\pi}{16} + \frac{1}{3}\left(2 +\sqrt\frac{2}{\pi} \Gamma\left(\frac{1}{4}\right)\Gamma\left(\frac{5}{4}\right)- \sqrt{2} \ {}_2F_1\left(-\frac{3}{4},\frac{1}{2};\frac{1}{4};-1\right)\right) \simeq 1.026 \,.
\end{equation}
The on-shell action includes a term which depends on the cutoff $\lambda_\infty$, but this term is the same regardless of the particular solution of the equations of motion (i.e.~it does not depend on the number of nodes and, hence, on the scale $\mu$). Since we are interested in comparing energies between different solutions with the same boundary conditions, we subtract \eqref{maxenergy} to the energy of a given embedding.
 
With this regularization the energy of the maximal embedding is clearly vanishing, whereas the energy of any other embedding is negative and monotonically increasing with the number of nodes. For the two parity-related minimal embeddings it reads 
\begin{equation}
E^0_F = -\frac{1}{V_3} (S_{D7}-S_{D7}^{max}) = - T_{D7} V_4 L^2 (b r^3_0 + a \mu^3) \simeq - N(g_s N) (b M_{KK}^3 + a M_{\mu}^3) ~,
\label{EF}
\end{equation}
where $a$ and $b$ are order one dimensionless constant and the energy scale $M_\mu$ is related to the length scale $\mu$ through the holographic radius/energy relation \cite{Peet:1998wn}, which we take here $\mu=M_\mu L^2/2$. This $\mathcal{O}(N)$ difference between the energy density of the maximal and the minimal embedding is related to the potential barrier that separates the degenerate vacua in field space.

One can consider a more general configuration, made of $F$ D7-branes. As already noticed, the up and the down embeddings are energetically equivalent. Hence, in the large $N$ limit in which flavor branes do not interact, one can choose $p$ of them being up and $F-p$ being down. As $p$ is varied from $0$ to $F$ all these configurations are energetically equivalent, with energy
\begin{equation}
\label{tot0pF}
E^0_{F,tot}=pE^0_F+(F-p)E^0_F=FE^0_F ~,
\end{equation}
which, indeed, does not depend on $p$.


\subsubsection{Massive case}
Let us now consider the inclusion of a bare quark mass. The quark mass can be viewed as a source for the meson operator $\bar\psi\psi$, which is described by the flavor brane profile, whose bending introduces the characteristic length scale $\mu$. We saw before that for large values of $\lambda$
\begin{equation}
u(\lambda) \sim \frac{1}{\sqrt{\lambda}} \,,
\end{equation}
so a small mass $m$ can be introduced by requiring that
\begin{equation}
\label{utom}
\lim_{\lambda\rightarrow\lambda_\infty} \sqrt{\frac{\lambda}{\mu}}u(\lambda) = 2\pi l^2_s m \,,
\end{equation}
which amounts to interpret the bare quark mass as the spatial separation between the D3 and the D7-branes along the common transverse direction $u$ in the ultraviolet regime of the theory. Indeed, quarks are the lightest modes of the open strings stretching between these branes, and get an energy proportional to their length.

We have seen that in the massless case there are two isoenergetic profiles for a flavor brane. Now we want to see if the inclusion of a small mass selects one of the two to be energetically favorite. This is to be expected, since the two different zero-node embeddings are not related anymore by a parity transformation $u\rightarrow -u$, because of the parity-breaking boundary condition at infinity. Intuitively, a small positive (negative) mass will make the up (down) embedding favorite, thus lifting the large $N$ vacuum degeneracy of the massless case. We now give a proof of this statement and, as a byproduct, we also derive the expression of the meson condensate for small quark mass.

Suppose to start from the massless case and perform a small change in the boundary condition of the flavor brane profile at the UV cutoff
\begin{equation}
\delta u_\infty = \sqrt{\frac{\mu}{\lambda_\infty}} 2\pi l^2_s \delta m \,.
\label{bndvalue}
\end{equation}
The corresponding variation of the on-shell action $S=\int d\lambda \cal L$ is given by
\begin{equation}
\delta S_{D7} =  \frac{\partial \cal L}{\partial \dot u} \delta u \bigg|^{\lambda=\lambda_\infty}_{\lambda=0} 
= - T_{D7} V_3 V_4 L^2 \left[ (r^4_0+4\rho^4)^{3/2} \frac{\lambda^4}{8\rho^8} \frac{\dot u ~ \delta u}{\sqrt{1+\dot u^2}} \right]^{\lambda=\lambda_\infty}_{\lambda=0} \,,
\end{equation}
where in the first step we have used the equation of motion. Since $\dot u(0)=0$, we get the following variation of the energy density
\begin{equation}
\label{cdef0}
\delta E_F = T_{D7} V_4 L^2 \lambda_\infty^2 \dot{u}(\lambda_\infty) \delta u_\infty \simeq \pm N \sqrt{g_s N} M^2_\mu ~ \delta m \,.
\end{equation}
Recall that $\dot u_\infty$ characterizes the massless embeddings and can have both signs. Now, if we give a positive (negative) mass, then the solution with $\dot u_\infty<0$ ($\dot u_\infty>0$) is preferred, i.e.~the up (down) embedding is selected. This means that the quark mass lifts the degeneracy between the two embeddings. Thus, for the energetically favorite embedding we have that (up to quadratic corrections in the quark mass)
\begin{equation}
E_F(m)=E_F(0) - c|m| \simeq E^0_F- N \sqrt{g_s N} M^2_\mu |m| 
\,.
\label{cdef}
\end{equation}
Note that this result implies that the fermion condensate is linear in $N$ and it is negative for positive mass and viceversa, since
\begin{equation}
\label{cdef2}
\braket{\bar\psi\psi} = \frac{dE_F}{dm} = -c ~ \mbox{sign}(m) \qquad \mbox{where}\qquad c\simeq N \sqrt{g_s N} M^2_\mu \ .
\end{equation}
It is now clear that the scale $\mu$ (or, equivalently, its energy counterpart $M_\mu$) is related to the scale of symmetry breaking. The fact that there is a discontinuity in the first derivative of the on-shell action (which maps to the free energy of the dual field theory) signals the presence of a first-order phase transition whenever one switches from an up to a down embedding or viceversa. This observation will play a crucial role later.

Let us now consider a configuration of $F$ flavor branes, with a common mass $m$. As discussed above, the degeneracy between up and down embeddings is lifted for $m \not =0$. Indeed, in the large $N$ limit where D7-branes do not interact, a configuration with $p$ flavor branes in the up embedding and $F-p$ in the down one would have a total energy
\begin{equation}
E_{F}(m,p)=p  (E^0_F - c m) +(F-p)  (E^0_F + c m)=FE^0_F + (F - 2p) cm ~.
\label{eftot}
\end{equation}
Clearly, if $m>0$ the minimal energy configuration occurs for $p=F$, whereas if $m<0$ for $p=0$. In both cases, the total energy will just be $F$ times $E_F(m)$, eq.~\eqref{cdef}. In the massless limit the degeneracy between up and down embeddings is regained, since the above equation reduces to eq.~\eqref{tot0pF}. This is insensitive to the value of $p$, and one recovers the degeneracy of all $F+1$ configurations obtained varying $p$ from 0 to $F$.


\section{Large $\boldsymbol N$ energetics of holographic QCD$_3$}
\label{energies}
In this section we want to derive the (large $N$) phase diagram of our holographic model. We will first discuss the generic structure of brane configurations describing its vacua. Then, using the results of the previous section, we will derive its full phase diagram and finally compare it with the pure QFT analysis. In section \ref{1/N} we will instead discuss how this is modified by taking into account $1/N$ corrections.

\subsection{Geometric structure of QCD$_3$ vacua}
\label{dualgeo}

In the previous section we have discussed, separately, embeddings of Chern-Simons and flavor branes in the D3-brane cigar geometry. Here we would like to consider configurations having both CS and flavor branes, since a vacuum of holographic QCD$_3$ would in general include both.

Actually, one cannot displace flavor branes at will, i.e.~independently of CS ones. Indeed, a CS/flavor brane configuration describing a vacuum of the theory should be compatible with UV data. The latter includes $N$, $F$, $m$ and the axion monodromy measured at the spacetime location holographically dual to the UV of the field theory. On the $(\lambda, u)$ plane (more precisely, it is the strip $0\leq\lambda\leq\lambda_\infty$), this is clearly the point $P_{UV}$ where all flavor branes intersect and the global symmetry is $U(F)$, i.e.~$P_{UV}=(\lambda_\infty,0)$ in the massless case.

We now show that in order to fix this monodromy to give a well-defined CS level $k \equiv k_b - F/2$, the number of CS branes must depend on $p$. The axion monodromy measures the effective CS level of the dual field theory as
\begin{equation}
\int_{S^1} F_1 = - k_{eff} \,,
\end{equation}  
where $S^1$ is a circle whose location in spacetime is specified, among all other coordinates, once we fix a point in the $(\lambda, u)$ plane. Since $C_0$ couples magnetically to D7-branes, the Bianchi identity of $F_1$ is violated by source terms which are delta functions picked at the location of both flavor and CS branes. As usual, this can be easily seen by computing the equations of motion of the dual form $C_8$, which instead couples electrically to D7-branes.

\begin{figure}[h]
\includegraphics[scale=0.4]{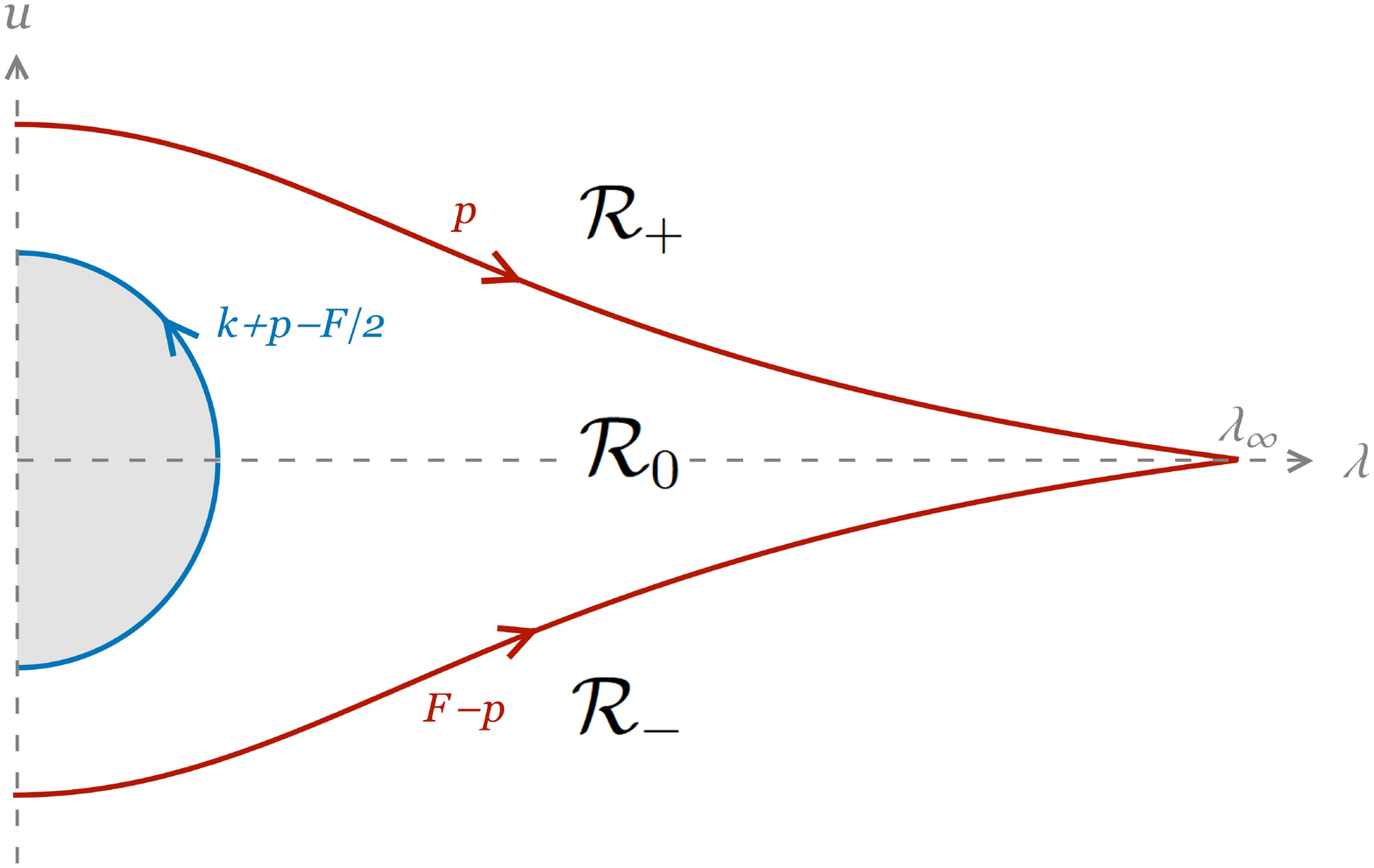}\centering
\centering
\caption{The configuration of flavor branes and (counterclockwise oriented) CS branes  in a (massless) $p$ sector. We interpret a negative number of counterclockwise branes as a positive number of clockwise branes.}
\label{psector}
\end{figure}

Let us define as `$p$ sector' (with $p=0,\dots,F$) a brane configuration with $p$ up branes (clockwise oriented), $F-p$ down branes (counterclockwise oriented) and $k_0$ counterclockwise oriented CS branes.\footnote{When $k_0<0$, the number of counterclockwise CS branes being $k_0$ actually means to have $|k_0|$ clockwise CS branes.} In order to determine $k_0$, we first compute $k_{eff}$, which is given by the following step function (we consider the massless case for definiteness)
\begin{equation}
k_{eff} = \begin{cases}
k_0-p  &\mbox{in } {\cal R}_+ \,, \\
k_0 &\mbox{in } {\cal R}_0 \,, \\
k_0 + F-p &\mbox{in } {\cal R}_- \,,
\end{cases}
\label{keffstep0}
\end{equation}
where ${\cal R}_+$, ${\cal R}_0$ and ${\cal R}_-$ are the regions in the $(\lambda, u)$ plane which are above, between and below flavor branes, respectively (see figure \ref{psector}). At the intersection point $P_{UV}$ both flavor branes count one-half and thus $k_{eff}=k_0 - p + F/2$ there. In order to have $k_{eff}=k$ at $P_{UV}$ we fix the number of counterclockwise CS branes to be
\begin{equation}
k_0=k+p-\frac{F}{2} \,.
\label{nCS}
\end{equation}
As a result, we can rewrite
\begin{equation}
k_{eff} = \begin{cases}
k-\frac{F}{2}  &\mbox{in } {\cal R}_+ \,, \\
k+p-\frac{F}{2} &\mbox{in } {\cal R}_0 \,, \\
k + \frac{F}{2} &\mbox{in } {\cal R}_- \,.
\end{cases}
\label{keffstep}
\end{equation}

\vspace{5mm}
This has a simple field theory interpretation. In region ${\cal R}_+$ $({\cal R}_-)$ it is as if all flavors have been integrated out with a negative (positive) mass. The effective CS levels read $k-F/2$ and $k+F/2$, respectively, consistently with $k_{eff}=k+\mbox{sign}(m) F/2$, as expected from field theory. In region ${\cal R}_0$ it is as if $p$ flavors have been integrated out with a positive mass and $F-p$ with a negative one, and hence the effective level is $k+p-F/2$.
In absence of flavor branes, when isotropy on the $(\lambda,u)$ plane is recovered, the effective CS level coincides with the bare level $k_b$ everywhere, as computed in eq.~\eqref{rrpurecs}.
Since $k$ is the time reversal odd CS level, we will take it to be non-negative without loss of generality.

Note that with the above argument we have recovered the number of CS branes $k_0$ in each $p$ sector that we argued to be there with the mechanism of $F-p$ up branes `pulled down' and wrapping the $S^5$ before sitting in the down embedding, described in section \ref{intro}. These topological arguments, and in particular eq.~\eqref{nCS}, hold regardless of flavor branes being massless or massive, the only difference being that $P_{UV}=(\lambda_\infty,u_\infty)$ in the massive case. 

In figure \ref{psector} the structure of a $p$ sector is depicted. Its low-energy dynamics is as follows.

\begin{itemize}
\item Flavor branes break spontaneously the gauge $U(F)$ symmetry (associated to the $F$ coincident branes in the UV) to $U(p)\times U(F-p)$. This  happens since the $F$ branes are spatially separated in the $u$ direction, as soon as we move towards the bulk. By Higgs mechanism the gauge bosons corresponding to the broken part of the gauge group become massive. These correspond to the $2p(F-p)$ lightest modes of the open strings having one extremum on one down brane and the other on one up brane. Instead, the gauge bosons corresponding to the up/up and the down/down open strings are still massless, signaling the presence of an unbroken $U(p)\times U(F-p)$ gauge group. The longitudinal components of the massive gauge bosons are holographically associated with Goldstone bosons in the dual field theory, through massless poles which must appear in correlators involving currents.
The global symmetry-breaking pattern is $U(F)\rightarrow U(p)\times U(F-p)$, leading to a number of Goldstone bosons which is indeed $F^2-p^2-(F-p)^2=2p(F-p)$. They parameterize a $\sigma$-model whose target space is
\begin{equation}
\mbox{Gr}(p,F)=\frac{U(F)}{U(p)\times U(F-p)} \ .
\end{equation}
Note that the RR five-form flux induces on the flavor D7-branes a term that should match the level $N$ Wess-Zumino term in the $\sigma$-model.

\item The CS branes give, at low energy, a three-dimensional $U(k+p-F/2)_{-N}$ theory (if $k+p-F/2>0$) or a $U(-k-p+F/2)_{N}$ theory (if $k+p-F/2<0$). These are pure three-dimensional Chern-Simons theories, since the Yang-Mills sector decouples (gluons get a large tree-level mass and decouple well before the theory reaches strong coupling). In both cases, these theories are level/rank dual to $SU(N)_{k+p-F/2}$.    
\end{itemize}
Thus, the IR dynamics of a $p$ sector is described by
\begin{equation}
\label{psectorIR}
\mbox{Gr}(p,F) \times SU(N)_{k+p-F/2} \,.
\end{equation}
The Grassmannian and the topological field theory are mutually decoupled, since the branes do not interact at leading order in the large $N$ expansion. 

Finally, let us observe that the $F+1$ sectors in \eqref{psectorIR} are the same which were found with QFT techniques in \cite{Armoni:2019lgb}. However, it is worth noticing that in the field theory analysis only the Grassmannian of each sector was derived from the effective potential of the theory, whereas the topological part was conjectured to be there (with minimal assumptions). In our construction both appear naturally in a simple geometrical way.

\subsection{Phase diagram of holographic QCD$_3$}
\label{phasediag}

The $F+1$ sectors discussed above are all possible configurations that can describe holographic QCD$_3$ vacua. We now want to uncover the phase diagram of the theory as a function of the fermion mass $m$, by minimizing the (free) energy over $p$.

At large $N$, we have the contribution from the flavor branes, the CS branes and the mass deformation, neglecting any kind of brane interactions. We have already written in \eqref{eftot} the contribution from the flavor branes for each $p$ sector. The CS contribution is just given by the number of CS branes in each sector times the energy density $E_{CS}$ of each of them, eq.~\eqref{ECS}. The total (free) energy of the $p$ sector is hence given by
\begin{equation}
E(p)=FE^0_{F} - 2cmp + Fcm + \left| k+p-\frac{F}{2} \right| E_{CS} \,,
\label{totalfree}
\end{equation}
where $E^0_{F}$, $c$ and $E_{CS}$ are all of order $N$. This formula is invariant under the transformation $p \rightarrow F-p$, $m\rightarrow -m$ and $k\rightarrow -k$, correctly implementing a time reversal transformation. We now distinguish different cases (recall that we can take $k\geq 0$ and that $0\leq p \leq F$). Let us define $m^*\equiv E_{CS}/(2c)$ and neglect the irrelevant constant shift $FE^0_{F} + Fcm$. 

\begin{enumerate}
\item If $k \geq F/2$, the quantity inside the absolute value is positive $\forall p$. So we have
\begin{equation}
E(p)=(E_{CS}-2cm)p + \left( k-\frac{F}{2} \right) E_{CS} \,,
\end{equation}
whose minimum is for $p=0$ if $m<m^*$ and for $p=F$ if $m>m^*$. If $m=m^*$ all $F+1$ vacua are degenerate.
The energy of the true vacuum as a function of $m$ hence reads
\begin{equation}
E_{vac} (m) = 
  \begin{cases}
      \left(k-\frac{F}{2}\right)E_{CS} & \mbox{if } m<m^* \,, \\
      -2cmF+\left(k+\frac{F}{2}\right)E_{CS} & \mbox{if } m>m^* \,.
    \end{cases}
\end{equation}
Since the derivative with respect to $m$ is discontinuous at $m=m^*$, the phase transition is first order. The vacuum $p=0$ is the pure TFT phase $SU(N)_{k-F/2}$ and the vacuum $p=F$ is the pure TFT phase $SU(N)_{k+F/2}$, where the global symmetry of the UV theory is unbroken. The resulting phase diagram, depicted in figure \ref{ps1}, is the same as the one in \cite{Aharony:2015mjs}, but with the phase transition being first order.
\begin{figure}[h]
\includegraphics[scale=0.55]{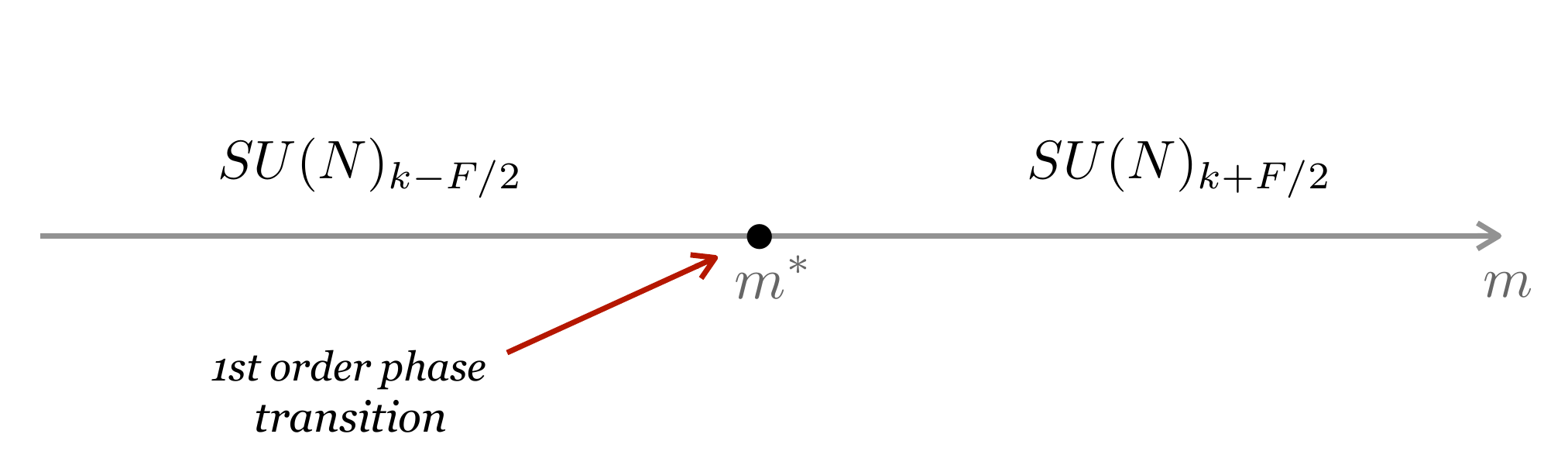}\centering
\centering
\caption{The phase diagram for $k\geq F/2$. At $m=m^*$ all $F+1$ vacua are degenerate.}
\label{ps1}
\end{figure}

\item If $k < F/2$, we have to see whether the quantity inside the absolute value is positive or negative. So we have to distinguish two subcases.

\begin{enumerate}
\item If $0 \leq F/2-k \leq p \leq F$
\begin{equation}
E(p)=(E_{CS}-2cm)p + \left( k-\frac{F}{2} \right) E_{CS} \,,
\end{equation}
whose minimum is for $p=F/2-k$ if $m<m^*$ and for $p=F$ if $m>m^*$. If $m=m^*$ the vacua $p=F/2-k,...,F$ are degenerate. Note that
\begin{equation}
\begin{split}
&E(F/2-k) =-2cm\left( \frac{F}{2} - k\right) \,, \\
&E(F) =-2cmF + \left( \frac{F}{2} + k\right)E_{CS} \,. \\
\end{split}
\end{equation}

\item If $0 \leq p \leq F/2-k \leq F$, then
\begin{equation}
E(p)=(-E_{CS}-2cm)p - \left( k-\frac{F}{2} \right) E_{CS} \,,
\end{equation}
whose minimum is $p=0$ if $m<-m^*$ and $p=F/2-k$ if $m>-m^*$, whereas if $m=-m^*$ the vacua $p=0,...,F/2-k$ are degenerate. Note that
\begin{equation}
\begin{split}
&E(0) = \left(\frac{F}{2} -k \right) E_{CS} \,, \\
&E(F/2-k) =-2cm\left( \frac{F}{2} - k\right) \,. \\
\end{split}
\end{equation}
\end{enumerate}

Looking at the different energies it follows that: if $m<-m^*$ the true vacuum is $p=0$, if $m=-m^*$ all the vacua $p=0,...,F/2-k$ are degenerate, if $-m^*<m<m^*$ the true vacuum is $p=F/2-k$, if $m=m^*$ all the vacua $p=F/2-k,...,F$ are degenerate, if $m>m^*$ the true vacuum is $p=F$.
The energy of the true vacuum as a function of $m$ hence reads
\begin{equation}
E_{vac} (m) = 
  \begin{cases}
      	\left(\frac{F}{2} -k \right) E_{CS} & \mbox{if } m<-m^* \,, \\
	-2cm\left( \frac{F}{2} - k\right) & \mbox{if } -m^*<m<m^* \,, \\
	-2cmF + \left( \frac{F}{2} + k\right)E_{CS} & \mbox{if } m>m^* \,. \\
    \end{cases}
\end{equation}
Since the derivative with respect to $m$ is discontinuous at $m=-m^*$ and $m=m^*$, the phase transitions are again first order. The vacua $p=0$ and $p=F$ are the same as before. The vacuum $p=F/2-k$ is described by the Grassmannian $\mbox{Gr}(F/2-k,F)$ with no TFT sector, and the symmetry-breaking pattern $U(F)\rightarrow U(F/2-k)\times U(F/2+k)$ takes place. The resulting phase diagram, depicted in figure \ref{ps2}, is analogous to the one discussed in \cite{Komargodski:2017keh}. The two first-order phase transitions take place at opposite values of $m$. Hence, at leading order in the large $N$ expansion, the point $m=0$ sits always inside the so-called quantum phase $\forall k < F/2$. For $k=0$ this is mandatory, as dictated by the Vafa-Witten theorem \cite{Vafa:1984xg,Vafa:1984xh,Vafa:1983tf}. Moreover, the width of the quantum phase in parameter space is given by 
\begin{equation}
\label{mcrit}
2 m^*=\frac{E_{CS}}{c} 
 \sim \frac{r_0^3}{\mu^2 l_s^2} \sim \sqrt{g_s N} \frac{M^3_{KK}}{M^2_\mu} ~,
\end{equation}
which is $\mathcal{O}(N^0)$ in the large $N$ expansion.  
\begin{figure}[h]
\includegraphics[scale=0.55]{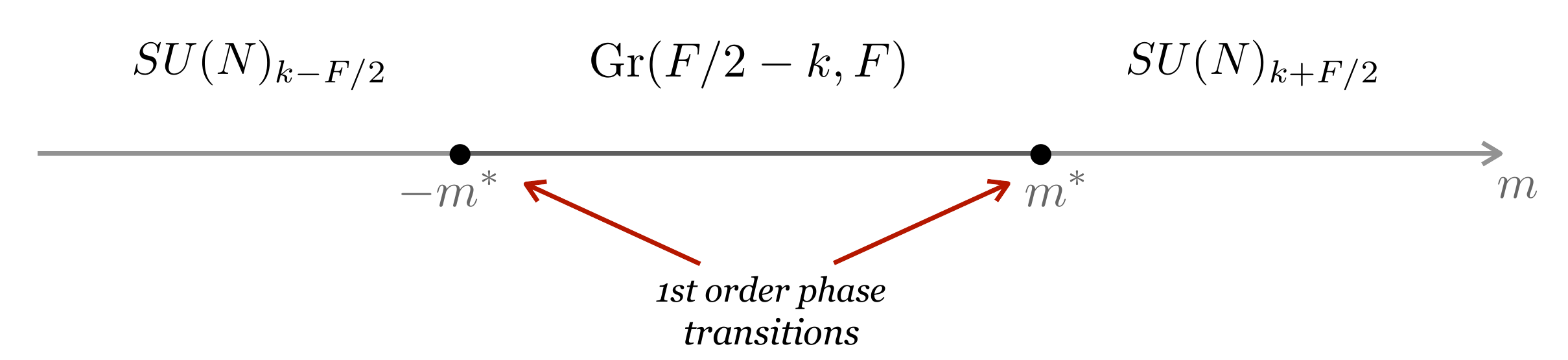}\centering
\centering
\caption{The phase diagram for $k< F/2$. At $m=-m^*$ the vacua with $p=0,\dots,F/2-k$ are degenerate, at $m=m^*$ the vacua with $p=F/2-k,\dots, F$ are degenerate.}
\label{ps2}
\end{figure}

When translated in the asymptotic boundary condition for the embedding through \eqref{utom}, the critical value of the mass corresponds to
\begin{equation}
u^* = \sqrt{\frac{\mu}{\lambda_\infty}} 2\pi l^2_s m^* \sim 
\sqrt{\frac{r_0}{\lambda_\infty}} \left( \frac{r_0}{\mu} \right)^{3/2} r_0 \ll r_0\ ,
\label{smalldisplacement}
\end{equation}
so the brane embedding at the critical value is still very close to the massless one.
Hence we are still well within the regime of small deviations from the latter and, as a consequence, also the Taylor expansion (\ref{cdef0}) is justified. One can consider next-to-leading order corrections to $m^*$, by computing the full mass dependence of the flavor brane energy. This gives a correction to \eqref{mcrit}, but clearly does not spoil the existence of first-order phase transitions and of a quantum phase whose width is $\mathcal{O}(N^0)$.
\end{enumerate}
As opposite to what happens in the large $N$ field theory description, in the holographic picture the quantum phase emerging for $k<F/2$ has a non-vanishing width already at leading order in the large $N$ expansion. This implies that the phase diagram displays a different structure in the two regimes $k \geq F/2$ and $k<F/2$. Interestingly, our phase diagram is identical to the one conjectured for the same theory, but at finite $N$ \cite{Komargodski:2017keh}.

The apparent discrepancy between holography and large $N$ field theory can be understood by pure field theory arguments, just recalling that our holographic set-up describes in fact a four-dimensional gauge theory compactified on a (supersymmetry breaking) circle. This reduces to a pure three-dimensional theory only in the limit where the radius of the compactified dimension is sent all the way to zero, equivalently $M_{KK} \rightarrow \infty$. As we will show below, in such limit our results reconcile with the pure 3d analysis.

Let us first notice that taking $E_{CS} = 0$, the two different phase diagrams displayed in figures \ref{ps1} and \ref{ps2} merge, both enjoying a single first-order phase transition at $m^*=0$, where all vacua are degenerate. This scenario is exactly the one proposed in \cite{Armoni:2019lgb} at the leading order in the large $N$ expansion. In particular, eq.~\eqref{totalfree} reduces to \eqref{eftot}, which exactly matches the effective potential computed with QFT techniques.

Large $N$ QCD$_3$ with massless probe quarks has only one scale, which is $\Lambda_3 = g^2_{\text{YM}_3} N$. All other quantities, such as the QCD-string tension (computed from the Wilson loop) and the fermion condensate, depend on $\Lambda_3$ in a way uniquely fixed by dimensional analysis (in particular both $\sigma$ and $\braket{\bar\psi \psi}$ scale as $\Lambda_3^{\, 2}$).

On the contrary, the four-dimensional theory our holographic model describes is a (S)YM$_4$ theory compactified on a circle, which is characterized by two parameters: the dimensionless 't Hooft coupling $\lambda_t=g^2_{\text{YM}_4}N \sim g_s N$ and the circle radius $1/M_{KK}$. In units of $M_{KK}$, different physical quantities depend on different powers of $\lambda_t$. This is a common feature of several holographic theories realized through compactification on $S^1$ of a higher-dimensional gauge theory. 

First of all recall that $M_{KK}$ sets the scale of the supersymmetry breaking masses of the fermions (and subsequently of the scalars). Hence, from the point of view of the 4d theory, we can assume that for energies above $M_{KK}$ the 't~Hooft coupling  is given by $\lambda_t \sim g_sN$ and it does not run, while at energies below $M_{KK}$ it runs as in pure YM (since flavors are quenched), with a dynamical scale defined by
\begin{equation}
\Lambda_4 = M_{KK} e^{-\frac{1}{\beta\lambda_t}}\ ,
\end{equation}
with $\beta$ an unimportant ${\cal O}(1)$ numerical positive factor.
Note that the relation above implies that at the compactification scale $M_{KK}$ the 4d theory is always in the deconfined phase, though for large $\lambda_t$ very close to the confining scale $\Lambda_4$.

At energy scales below $M_{KK}$ the theory becomes effectively three-dimensional. Hence what is now relevant is the 3d dynamical scale. We first identify
\begin{equation}\label{3dcoupl}
g^2_{\mathrm{YM}_3} \sim g^2_{\mathrm{YM}_4}M_{KK} \ .
\end{equation} 
The above relation must be understood at the matching scale, i.e.~at $E\sim M_{KK}$. The 3d dynamical scale is then 
\begin{equation}\label{3dlambda}
\Lambda_3 \sim g_s N M_{KK} \sim \lambda_t M_{KK} \ .
\end{equation}
It is now obvious that the limits $\lambda_t\to 0$ and $\lambda_t\to \infty$ describe very different regimes. For $\lambda_t\to \infty$, the compactification scale is very close to the confining scale from the 4d point of view. Below that scale, from the 3d point of view one is already deeply in the confining regime. Thus one is never really in a 3d theory with perturbative degrees of freedom. When $\lambda_t\to0$, instead, the theory compactifies when it is still in the perturbative regime, both in 4d and also in 3d. Hence the evolution can go on towards the IR, until the theory confines as a purely 3d theory. We will call the `3d limit' the latter, when one sends $M_{KK}\to \infty$ holding the 3d scale fixed. 

From the 4d perspective, the non-vanishing CS level is obtained by turning on an $x^3$-dependent $\theta$ angle. Concretely, this is implemented as in eqs.~\eqref{C0coupl}--\eqref{C0CS}. This would produce $k$ equally spaced domain walls (more precisely, interfaces), i.e.~each time $\theta=\pi \mod 2\pi$.\footnote{Indeed, the D7-branes that engineer the CS level in the present set-up are straightforwardly related by T-duality to the D6-branes that holographically engineer the $\theta = \pi$ domain walls of YM$_4$ \cite{Witten:1998uka}.} Deforming the varying $\theta$ to a step function can bring all domain walls together, and produce a level $k$ CS term for the $SU(N)$ gauge field, when reduced to the domain wall (see \cite{Gaiotto:2017yup,Gaiotto:2017tne,Kan:2019rsz}). 
The tension of such domain walls is given by $T_{DW}\sim N\Lambda_4^{\, 3}$. In the 3d limit, this becomes
\begin{equation}\label{tdw3d}
T_{DW}\sim N \Lambda_3^{\, 3} \frac{1}{\lambda_t^3}e^{-\frac{3}{\beta\lambda_t}}\to 0 \ .
\end{equation}
We thus see that domain walls (which correspond to CS branes in our set-up) become tensionless in the 3d limit, so that the CS level becomes a feature of the 3d theory and is no longer associated to an object that has been added to the theory. Hence, it does not come as a surprise that as the energy of CS branes vanishes, $E_{CS} \rightarrow 0$, our phase diagram becomes identical to the field theory one. In fact, our result can be regarded as an independent check for the validity of the analysis performed in \cite{Armoni:2019lgb}. 

The consistency of this picture can be understood also from the point of view of the large $N$ expansion. The finite width of the quantum phase in figure \ref{ps2} is proportional to $m^*$ and thus to $E_{CS}$, suggesting that in the four-dimensional theory compactified on a circle the large $N$ expansion breaks down. This is actually the case and has a clear field theory origin. As we just emphasized, the 3d Chern-Simons term is implemented through a varying $\theta$ angle in the parent 4d theory. This generates $k$ interfaces, described in the holographic set-up by wrapped D7-branes. These objects have tension proportional to $N$ and this indeed spoils the large $N$ counting rules of the four-dimensional theory. In the 3d limit, where $E_{CS}\rightarrow 0$, the consistency of large $N$ counting is recovered.\footnote{We thank Zohar Komargodski for a discussion on this point.}

\section{$\boldsymbol{1/N}$ corrections}
\label{1/N}

All what we have been discussing so far was at leading order in the large $N$ expansion. Here we would like to consider the first next-to-leading order corrections. 
What we have to do is to compute $1/N$ contributions to the free energy and to minimize such contributions over the different $p$ sectors. Since we are interested in the vacuum energy, we can safely neglect the contribution of the gauge fields on the probe branes, whose fluctuations describe instead the dynamical degrees of freedom of the theory.

Recall that at leading order we considered the sum of all contributions coming from the tensions of the probe branes, i.e.~the DBI part of the on-shell action. Clearly, the presence of a Ramond-Ramond axion introduces a term in the action given by
\begin{equation}
S_{RR} = - \frac{1}{2(2\pi)^7 l_s^8} \int d^{10}x \sqrt{-g} |F_1|^2 \,,
\label{srr}
\end{equation}
where the integral has to be performed over the entire spacetime. This contribution is $1/N$ suppressed with respect to the DBI, since it has no explicit factor of $g_s$.

Given the result in \eqref{keffstep}, the only $p$-dependent part of the on-shell action is given when performing the integral in \eqref{srr} over the region ${\cal R}_0$, where $k_{eff}=k_0$. Thus, neglecting $p$-independent terms we get
\begin{equation}
S_{RR}= - \frac{1}{2(2\pi)^7 l_s^8} V_3 V_4 L^2 \int_{{\cal R}_0} d\lambda \ du \int_{S^1} dx^3 \ r^3 \ |F_1|^2 \,,
\end{equation}
where $r$ is expressed in terms of the radial coordinate $\rho^2=u^2+\lambda^2$ as in eq.~\eqref{change}. We can extract the $1/N$ correction to the free energy density as
\begin{equation}
E_{1/N}(p) = -\frac{S_{RR}}{V_3} = \Delta \left(k+p-\frac{F}{2}\right)^2 \,,
\label{total1n}
\end{equation}
where $\Delta$ is a positive constant given by
\begin{equation}
\Delta \sim (g_s N)^2 M_{KK} M_\mu^2 \,.
\end{equation}
As we are going to show below, the positiveness of $\Delta$ has important implications on the phase diagram. The corresponding field theory quantity was argued to be positive in \cite{Armoni:2019lgb} using consistency with the Vafa-Witten theorem. In our holographic context, instead, we cannot use a similar argument, since the phase $p=F/2$ in the $k=0$ massless case is already selected at leading order in large $N$ and the Vafa-Witten theorem is surely satisfied regardless the sign of $\Delta$.\footnote{Moreover, stricly speaking, the Vafa-Witten theorem does not necessarily apply to the gauge theory realized by our D-brane set-up, due to the presence of Yukawa couplings.} Nicely, our geometric set-up encodes in a simple way the $1/N$ corrections and allows to determine the value of $\Delta$ in terms of the defining parameters of the model, besides showing its positivity. Similar arguments as the ones presented here were used in the Sakai-Sugimoto model to compute the Witten-Veneziano mass of the $\eta'$ meson \cite{Sakai:2004cn}.

In our holographic set-up, having a positive $\Delta$ implies that at the microscopic level locally parallel probe branes effectively repel each other. The fact that the contribution of $\Delta$ to the free energy is a $1/N$ effect is consistent with brane interactions being next-to-leading order in $g_s$. This effective repulsion can be rephrased by saying that in our non-supersymmetric set-up, the effective tension of the probe branes is smaller than their effective charge. A similar effect was found in a different non-supersymmetric set-up, which allowed to perform such computations \cite{Bonnefoy:2018tcp}.

We can now sum the $1/N$ contribution in \eqref{total1n} to the leading order one in \eqref{totalfree}. Neglecting again $p$-independent contributions, one gets for the total energy 
\begin{align}
\label{energy0}
E= \left| k + p -\frac{F}{2} \right| E_{CS}-2 \,c \,m' p + \Delta \, p (p-F)~,
\end{align}
where $m' = m - k \Delta/c$ accounts for the expected ${\cal O}(1/N)$ shift of the fermion mass due to a non-vanishing CS level $k$.
The important point is that the extremization problem is modified comparing to the leading order one by the addition of a subleading quadratic term, proportional to $\Delta$. As we will see, the final phase diagram crucially depends on such quantity.

\subsection{Phase diagram}
We now establish the phase diagrams as the common flavor mass $m$ is varied. Recall that at leading order (no $p^2$ term) we found one first-order transition at $m^*=E_{CS}/(2c)$ when $k \geq F/2$ and two first-order transitions at $\pm m^*$ when $k< F/2$. 

\begin{itemize}
\item $k \geq F/2$\\
In this case $ k + p - F/2 >0$ so the expression \eqref{energy0} becomes 
\begin{equation}
\label{energy1}
E= \Delta p^2 + p \left(E_{CS} - 2 \, m' c - \Delta F \right)~, 
\end{equation}
which we need to minimize as a function of $p$. Given that $\Delta>0$ the function has a local minimum at
\begin{equation}
p_{min} = - \frac{E_{CS}- 2 \,m' c - \Delta F}{2 \Delta}~.
\end{equation}
The value of $m'$ such that $p_{min} = p+\frac{1}{2}$ gives the mass for which a phase transition occurs between the phases labeled by $p$ and $p+1$. A straightforward computation gives for the (shifted) mass the value 
\begin{equation}\label{Fpts}
m_p= m^* + \frac{\Delta}{2 c} (2p-F+1)~.
\end{equation}
As a check one can see that for the above value of the mass the vacua labeled by $p$ and $p+1$ are degenerate in energy while all others have higher energy. 

This analysis holds for any of the $F+1$ values of $p$, so we get $F$ first-order phase transitions at values $m_p$ defined by \eqref{Fpts}. These are ${\cal O}(1/N)$ away from the leading order value $m^*$, at which all $F+1$ vacua were degenerate at large $N$. The resulting phase diagram is depicted in figure \ref{ps3}.
\begin{figure}[!htbp]
	\includegraphics[scale=0.6]{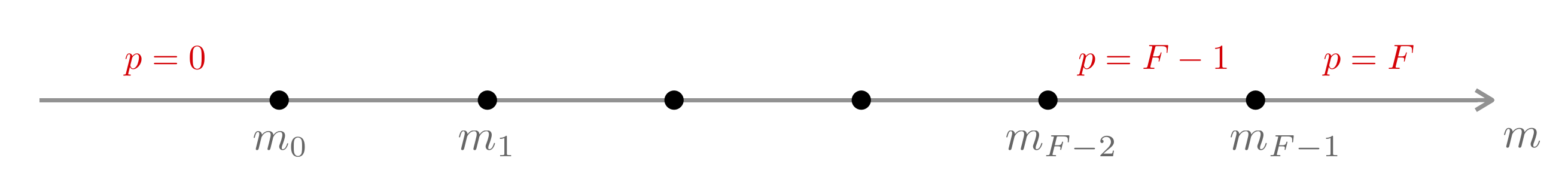}\centering
	\centering
	\caption{Phase diagram for $k\geq F/2$ at order $1/N$. All intermediate phases have ${\cal O}(1/N)$ widths. All masses $m_p$ are positive. At each critical point a first-order phase transition occurs where two phases become degenerate.}
\label{ps3}
\end{figure}

\item $k<F/2$\\
In this case, the sign of $k + p - F/2$ is not fixed. We have two expressions for the energy, eq.~\eqref{energy1} for $p \geq F/2 - k$ and 
\begin{equation}
\label{energy2}
E= \Delta p^2 - p \left(E_{CS} + 2 \, m' c + \Delta F \right)~, 
\end{equation}
for $p < F/2-k$. One can run the same argument as before and find the value of the mass for which a phase transition occurs between nearby phases. This is the expression  \eqref{Fpts} for $p \geq F/2 - k$  and 
\begin{equation}
m_{p}= - m^*+ \frac{\Delta}{2 c} (2p-F+1)~.
\end{equation}
for $p < F/2 - k$. All in all we get again $F$ phase transitions and a set of intermediate phases whose widths are ${\cal O}(1/N)$ suppressed but the one described by $p=F/2-k$, the quantum phase already present at leading order in the large $N$ expansion. The corresponding phase diagram is reported in figure \ref{ps4}.
\begin{figure}[!htbp]
	\includegraphics[scale=0.58]{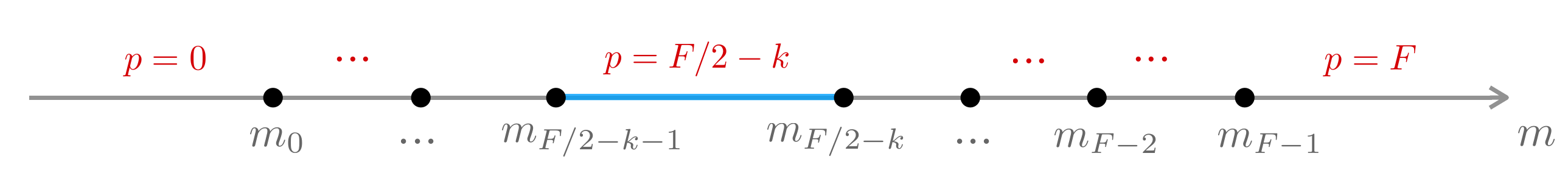}\centering
	\centering
	\caption{Phase diagram for $k <  F/2$ at order $1/N$. The phase $p=F/2-k$, in blue, has  ${\cal O}(1)$ width and it is the quantum phase already present at large $N$. All others intermediate phases have ${\cal O}(1/N)$ width. Masses on the right (left) of the quantum phase are positive (negative). At critical points the corresponding adjacent phases become degenerate.}
	\label{ps4}
\end{figure}
\end{itemize}

The final result we get, figures \ref{ps3} and \ref{ps4}, is a phase diagram similar to the one obtained in \cite{Armoni:2019lgb} by field theory arguments at order $1/N$, the only difference being that the purely quantum phase $p=F/2-k$ in the holographic set-up has an ${\cal O}(1)$ width. As for the leading order result, see the discussion in section \ref{phasediag}, this originates from $m^* \propto E_{CS}$ not being zero, which is a property of the four-dimensional theory our holographic set-up describes, and which vanishes in the strict 3d limit. 

\section{Bosonization dualities from string theory}
\label{bosondual}

The analysis of the low-energy properties of QCD$_3$ led to propose new infrared dualities between QCD$_3$ and three-dimensional gauge theories coupled to fundamental scalars. Motivated by explicit computations in the limit where both $N$ and $k$ are large, it was proposed in \cite{Aharony:2015mjs} that, for any $k\geq F/2$, QCD$_3$ has a single IR fixed point which can be equivalently described by a bosonic theory
\begin{equation}
SU(N)_k+F~\psi \qquad \longleftrightarrow \qquad U(k+F/2)_{-N}+F~\phi \,.
\label{k>f2}
\end{equation} 
This proposal was extended in \cite{Komargodski:2017keh} to the case $k<F/2$, where QCD$_3$ develops two IR fixed points (if the corresponding transitions are assumed to be second-order), each being described by a different dual bosonic theory. One of them is the same as in \eqref{k>f2}, while the other is its `time reversal' version, giving rise to a second boson/fermion duality
\begin{equation}
SU(N)_k+F~\psi \qquad \longleftrightarrow \qquad U(F/2-k)_{N}+F~\phi \,.
\label{k<f2}
\end{equation} 
Assuming that the dynamics prefers to maximally Higgs the gauge group on the bosonic side (which can be proven under the hypothesis of a quartic scalar potential, see \cite{Argurio:2019tvw,Armoni:2019lgb}), the phases of the bosonic theories for positive and negative squared masses agree with the ones of QCD$_3$.

Boson/fermion dualities are fully meaningful in presence of an IR fixed point, i.e.~when at the second-order phase transition the same conformal field theory emerges. In the case of first-order transitions, dualities are less powerful but still carry non-trivial information. In particular, dual theories have the same vacua and the same phases under relevant deformations. For large $N$ QCD$_3$ this has been analyzed in \cite{Armoni:2019lgb}, where the vacuum structure with $F+1$ degenerate vacua was matched with the same bosonic theories as in \eqref{k>f2} and \eqref{k<f2}, equipped with a suitably chosen sextic scalar potential. This has been recently extended to the case where the UV flavor symmetry is explicitly broken \cite{Baumgartner:2020rum}.

As we are going to argue below, the holographic picture furnishes a simple geometric understanding on how such dual bosonic theories arise, similarly as it happens in the holographic description of QCD$_4$ domain walls \cite{Argurio:2018uup}.

Let us first consider the case $k\geq F/2$, when the $|k+p-F/2|$ CS branes in each $p$ sector are always counterclockwise. Take the situation where this number is maximal, so that $p=F$ and all flavor branes are up. The lightest modes of the open strings stretching between CS D7-branes and flavor ones are scalars with one gauge index and one flavor index, i.e.~$F$ fundamental scalars of $U(k+F/2)$. The CS/CS open strings provide the gauge sector of the theory, while the masslessness of the up/up open strings signals the presence of an unbroken $U(F)$ global symmetry. Being the CS branes counterclockwise, the level of the theory is $-N$. All in all, we have the theory on the right-hand side of \eqref{k>f2}. All other configurations correpond to a partially Higgsed gauge group $U(k+p-F/2)$ and to a $U(F)$ global symmetry spontaneously broken to $U(p) \times U(F-p)$. The critical distance between the various brane embeddings should translate (though in a possibly complicated way) into the parameters of the scalar potential which guarantee such a vacuum structure.

In the case $k< F/2$, a single dual bosonic theory is not sufficient to describe all vacua, since the CS branes change orientation at $p=F/2-k$. It is easy to realize that the same bosonic theory described above includes all phases with $p=F,...,F/2-k$. To characterize all other phases, take the configuration where the number of clockwise CS branes is maximal, i.e.~when $p=0$ and all flavor branes are down. The gauge group is now $U(F/2-k)$ at level $+N$ and the flavor symmetry $U(F)$ is unbroken. All the phases with $p=F/2-k,...,0$ are described by moving up one by one the flavor branes, until one reaches the geometric configuration where there are no CS branes. This is described by the dual bosonic theory on the right hand-side of \eqref{k<f2}.

In our holographic picture, the fact that one of the two critical points stays at the same parametric value $m=m^*$, for both the $k\geq F/2$ and $k<F/2$ regimes, suggests that the same dual bosonic theory can describe the neighborhood of that critical point in both regimes. For $k<F/2$ a second critical point shows up, at exactly the time-reversed critical mass. This suggests that the dual bosonic description around $-m^*$ is indeed given by the time reversal of the dual bosonic description of the critical point $m^*$. This observation supports our considerations above.

Nicely, in the string theory picture the shift of the CS level due to the integration of massive fermions can be equivalently interpreted as Higgsing of the gauge group of the dual bosonic theories. Consequently, the field theory assumption of maximal Higgsing is mapped into the requirement that the preferred vacua (in the massless case) are the ones with the minimal number of CS branes. In our set-up this fact is not an assumption, since it easily follows from the minimization of the on-shell energy density on the gravity side, at leading order in the large $N$ expansion, as shown in section \ref{phasediag}. Moreover, the holographic picture makes manifest the necessity of two mutually non-local dual bosonic theories in the case $k<F/2$, and gives an indirect check that the vacua of QCD$_3$ can be captured by a dual bosonic description even in the absence of a proper IR fixed point. 

Let us finally comment on the scalar potential of the dual bosonic theory, as it emerges from the stringy description above. At leading order in $N$ this is given by the sum of single trace operators up to a sextic term (higher order terms being irrelevant). This potential should guarantee the vacuum structure we discussed in section \ref{phasediag}, including maximal Higgsing and the existence of first-order transitions. At subleading order in $N$ double trace operators have to be included in the potential. In particular, a double trace quartic operator gives a contribution which has the same form as the $\Delta$ contribution in eq.~\eqref{total1n}. Hence, it is natural to identify $\Delta$ (up to a positive dimensionful constant) with the coupling of the double trace quartic operator of the dual bosonic theory. It is indeed the sign of $\Delta$ which fixes the topology of the phase diagram, once $1/N$ corrections are included.

\section{Comments and outlook}
\label{conc}

In this concluding section, there are some aspects we would like to comment upon.

The first aspect regards the large $N$ expansion itself, see the discussion at the end of section \ref{phasediag}. While the discrepancy between our phase diagram and that of \cite{Armoni:2019lgb} disappears in the 3d limit, since $E_{CS}\rightarrow 0$, one cannot exclude that the holographic result contains more information than a mere contamination from the parent 4d theory. For instance, it is suggestive that a quantum phase, which is believed  to  exist in QCD$_3$ at finite $N$ \cite{Komargodski:2017keh}, naturally emerges in the holographic set-up already at leading order, giving a phase diagram which is in fact identical to the one conjectured to hold at finite $N$. In principle, it is not guaranteed that the large $N$ expansion strictly holds in CS QCD$_3$. In particular, the structure of the QCD$_3$ phase diagram at next-to-leading order in $1/N$ presented in \cite{Armoni:2019lgb} was obtained under the assumption that the large $N$ expansion works. Our holographic analysis provides some more evidence for the validity of this assumption, since the violating term is a pure 4d effect, but we believe it would be interesting to investigate this point further.

A second aspect we would like to comment upon regards the asymptotic solution of the equation of motion for the brane profile $u$ admitting complex roots, see section \ref{m=0}. This is because the field $u$ is below the Breitenlohner-Freedman (BF) bound \cite{Breitenlohner:1982jf}. This is not uncommon in holographic models and has interesting implications. In particular, as originally discussed in \cite{Kaplan:2009kr}, the violation of the BF bound  can be associated to loss of conformality in the dual field theory (see also \cite{Jensen:2010ga,Iqbal:2010eh}). This suggests a connection between the first-order nature of the phase transition in the large $N$, finite $k$ and $F$ regime we have investigated, and the nature of the scalar field $u$ in the background \eqref{bgD7f}. 

Seemingly, as $k$ is increased it is expected that the phase transition changes its nature and becomes second order for $k \sim N$ \cite{Aharony:2011jz,Giombi:2011kc,Aharony:2012nh,GurAri:2012is,Aharony:2012ns,Jain:2013py,Jain:2013gza}. In our model, the large $k$ regime can be investigated by backreacting the CS branes, which in the present paper were treated as probes instead. In a holographic model in which both the D3 and the CS branes are backreacted one should then expect the existence of a critical value $k_c=k_c(N)$ above which the roots for the characteristic polynomial for the (asymptotic) equation for $u$ turn real. This is a compelling scenario worth being investigated.

\section*{Acknowledgments} 
We thank Ofer Aharony, Marco Bill\`o, Aldo Cotrone, Thomas Dumitrescu, Guido Festuccia, Alberto Lerda and Javier Tarrio for discussions. We are grateful to Zohar Komargodski and Shigeki Sugimoto for discussions and useful comments on the manuscript.
R.A. and P.N. acknowledge support by IISN-Belgium (convention 4.4503.15) and by the F.R.S.-FNRS under the ``Excellence of Science" EOS be.h project n.~30820817, M.B. and F.M. by the MIUR PRIN Contract 2015 MP2CX4 ``Non-perturbative Aspects Of Gauge Theories And Strings" and by INFN Iniziativa Specifica ST\&FI. R.A. is a Research Director and P.N. is a Research Fellow of the F.R.S.-FNRS (Belgium).
The authors warmly thank each others' institutes for the kind hospitality during the preparation of this work.



\end{document}